\documentclass[fleqn,usenatbib]{mnras}

\usepackage[T1]{fontenc}
\usepackage{ae,aecompl}

\usepackage{graphicx}	
\usepackage{amsmath}	
\usepackage{amssymb}	

\usepackage[T1]{fontenc}
\usepackage{verbatim}
\usepackage{graphicx}
\usepackage{subfigure}
\usepackage{tabularx}
\usepackage{amsmath}
\usepackage{natbib}
\usepackage{xspace}
\usepackage{hyperref}
\graphicspath{{figures/}}

\makeatletter

\usepackage{txfonts}
\usepackage{cleveref}

\newcommand{\unit}[1]{\ensuremath{\, \mathrm{#1}}}
\newcommand{\athenaplus}{{\sc athena++}}          
\newcommand{\dispatch}{{\sc dispatch}}          
\newcommand{\ramses}{{\sc ramses}}          
\newcommand{\mesa}{{\sc mesa}\xspace}          
\newcommand{\radmc}{{\sc radmc3d}\xspace}          
\newcommand{\vapor}{{\sc vapor}}          

\newcommand{\Fig}[1]{Fig.~\ref{fig:#1}}    
\newcommand{\Figure}[1]{Figure~\ref{fig:#1}}
\newcommand{\Table}[1]{Table~\ref{tbl:#1}}

\newcommand{\Msunyr}{\mathrm{M}_{\sun}\,\unit{yr}^{-1}}

\title[Episodic accretion]{Episodic accretion: the interplay of infall and disc instabilities}

\author[M. Kuffmeier et al.]{
Michael Kuffmeier,$^{1}$\thanks{E-mail: kueffmeier@nbi.ku.dk (MK)}
S{\o}ren Frimann,$^{2}$
Sigurd S. Jensen$^{1}$
and Troels Haugb{\o}lle$^{1}$
\\
$^{1}$Centre for Star and Planet Formation, Niels Bohr Institute and Natural History Museum of Denmark, University of Copenhagen,\\
{\O}ster Voldgade 5-7, DK-1350 Copenhagen K, Denmark\\
$^{2}$Institut de Ci\`{e}ncies del Cosmos, Universitat de Barcelona, IEEC-UB, Mart\'{i} Franqu\`{e}s 1, E-08028 Barcelona, Spain
}

\date{Accepted XXX. Received YYY; in original form ZZZ}

\pubyear{2017}

\begin{document}
\label{firstpage}
\pagerange{\pageref{firstpage}--\pageref{lastpage}}
\maketitle

\begin{abstract}
Using zoom-simulations carried out with the adaptive mesh-refinement code \ramses{} with a
dynamic range of up to $2^{27}\approx 1.34 \times 10^{8}$ we investigate
the accretion profiles around six stars embedded in different
environments inside a (40 pc)$^3$ giant molecular cloud,
the role of mass infall and disc instabilities on the accretion profile,
and thus on the luminosity of the forming protostar.
Our results show that the environment in which the protostar is embedded
determines the overall accretion profile of the protostar.
Infall on to the circumstellar disc may trigger gravitational disc instabilities in the disc
at distances of around $\sim$10 to $\sim$50 au leading to rapid transport of angular momentum and
strong accretion bursts.
These bursts typically last for about $\sim$10 to a $\sim$100 yr,
consistent with typical orbital times at the location of the instability,
and enhance the luminosity of the protostar. Calculations with the stellar evolution
code \mesa\ show that the accretion bursts induce significant changes 
in the protostellar properties, such as the stellar temperature and radius.
We apply the obtained protostellar properties to produce synthetic observables with \radmc\
and predict that accretion bursts lead to observable enhancements around $20$ to $200$ $\mu$m
in the spectral energy distribution of Class 0 type young stellar objects.
\end{abstract}

\begin{keywords}
star formation -- protoplanetary disc formation -- adaptive mesh refinement
\end{keywords}

\section{Introduction}
Protostars form as a consequence of the gravitational collapse of prestellar cores and grow by accreting material from the surrounding environment. In the early stages of star formation the protostellar luminosity is predominantly determined by the mass accretion rate on to the protostar. In the simplest picture of star formation, corresponding to the collapse of a singular isothermal sphere in hydrostatic equilibrium \citep{Shu:1977ef}, the accretion rate on to the protostar is predicted to have a constant value of $2\times 10^{-6}\, \Msunyr$.
Similarly, an average accretion rate of $2\times 10^{-6}\, \Msunyr$ is needed to produce a one solar mass star during the expected 0.5\,Myr duration of the embedded phase of star formation \citep{2009ApJS..181..321E}. Assuming a typical mass of 0.5\,M$_{\sun}$ and a stellar radius of 2\,R$_{\sun}$ this corresponds to an accretion luminosity of $\sim$10\,L$_{\sun}$, which is significantly above the median luminosity of $\sim$1\,L$_{\sun}$ inferred from observations \citep{2009ApJS..181..321E, Kryukova:2012ea, Dunham:2013do}. The inability of simple physical arguments to reliably predict the typical luminosity inferred by observations is known as the 'luminosity problem' \citep{1990AJ.....99..869K}, and suggest that the accretion on to young stellar objects (YSOs) evolves over time. Possible solutions to the luminosity problem include a smooth decline in the accretion rate (possibly interspersed with variations from the large-scale infall) from early to late stages \citep{Offner:2011ex,2014ApJ...797...32P} as well as short episodic burst of high intensity likely driven by instabilities in the circumstellar disc \citep{2010ApJ...718L..58I,2010ApJ...723.1294V,2015ApJ...805..115V,2011ApJ...730...32S,2012MNRAS.427.1182S}.

Observationally, episodic accretion events are well-established in FU~Orionis (FUor) type objects, which are pre-main-sequence stars showing long-lived optical accretion bursts with luminosity enhancements of one or two magnitudes \citep{Herbig:1966jo,Herbig:1977gf}. Along with the shorter duration less intense EX~Lupin (EXor) type accretion bursts \citep{Herbig:1989wl} these objects constitute the main direct observational evidence of episodic accretion amongst the YSOs. Bursts are relatively rare and it is therefore difficult to accurately estimate the average interval between bursts. \citet{Scholz:2013ij} surveyed 4000 YSOs in the Galactic plane over two different epochs set 5 yr apart finding four burst candidates. Based on their analysis they argue for a typical burst interval between 5\, kyr and 50\,kyr.

For the less evolved Class~0 and~I protostars that are still embedded within their natal envelope it is challenging to detect episodic outbursts directly, as these objects are only detectable at mid-infrared wavelengths and beyond. To date, we are aware of three direct observations of accretion burst towards embedded protostellar systems, 
namely the cases of OO Serpentis \citep{2007A&A...470..211K}, HOPS~383, which increased its 24\,$\mu$m flux by a factor of 35 between 2004 and 2008 \citep{Safron:2015ew}, 
and the massive hot core MM1 in NGC63341 with an increase in luminosity by more than a factor of 50 \citep{2017ApJ...837L..29H}. For this reason indirect methods have to be employed, which include the detection of bullets in the outflows towards some protostars suggesting variations in the accretion rate with intervals between approximately 0.1 and 6\,kyr (e.g.\ \citealt{Bachiller:1990va,Bachiller:1991ui,Lee:2009gp,Arce:2013kt,Plunkett:2015in}). Another option is to use chemistry to trace past accretion bursts.
For example, the evaporation of CO ice into the gas phase during an accretion burst, which leads to CO line emission over a larger region, can be used as tracer of accretion bursts and remains detectable even after the burst has ended. Observations of this situation towards nearby low-mass star-forming regions suggest an average burst interval between 20 and 50\,kyr \citep{Jorgensen:2015kz,2017A&A...602A.120F}.

Until now, simulations with a high enough physical resolution to resolve disc dynamics have not accounted for infalling material from the interstellar medium, while simulations with a large enough box-size to account for the dynamics in Giant Molecular Clouds (GMCs) lacked the resolution to resolve discs. \citet{Liue1500875} investigated the process of episodic accretion by modelling the dynamics of four particular systems, where they added infalling material to the set-up in an ad hoc manner. However, stars are embedded in different environments, often being part of stellar clusters associated with filaments of gas, and non-trivial correlations between gas density, flow structure and magnetic fields. A consistent study taking into account the full dynamics in GMCs while simultaneously resolving the scales of protoplanetary discs remains the ultimate goal to constrain the dominant mechanism of episodic accretion.

In this paper, we overcome these limitations using very deep adaptive mesh refinement simulations. 
We investigate the accretion process of six
stars embedded in different environments on the basis of previously presented zoom-simulations of GMCs (\citet{2017ApJ...846....7K}.
In the zoom-simulations grid cells are concentrated around each individual star, but the
global GMC simulation is retained and evolved.
This allows us to investigate episodic accretion in different environments
maintaining the boundary conditions including a correct magnetic field anchoring and a realistic mass infall history.
Using this method we are able to simulate the full molecular cloud environment, while simultaneously resolving
au-scale and in one case sub-au scale dynamics in the accretion disc. 
To make synthetic observations we need an account of the total luminosity of our stars. This is done by computing
the detailed stellar evolution using the stellar structure code \mesa. We have modified {\mesa} to include the effects of time-dependent accretion
and heating by the accretion shock at the surface of the star. Using the stellar radius, effective temperature,
and thermal efficiency at the accretion shock we can compute total luminosities of the stars. The luminosities are used to investigate
the observational signatures in the dust continuum of the protostellar system
by carrying out radiative transfer simulations with the Monte Carlo code \radmc.

We present the zoom-simulations in section 2.1,
the post-processing with the stellar evolution \mesa\ in section 2.2,
and the setup for the synthetic observations of the dust continuum in section 2.3.
In section 3, we analyse the accretion rates obtained in the zoom-simulations,
the models of the stellar evolution, as well as how the protostellar systems
would appear if observed in the dust continuum.
We discuss our results and compare them to observations in section 4,
and draw the conclusions in section 5.
\section{Methods}
\subsection{Zoom-simulations with \ramses}
We briefly summarize the setup of the
magnetohydrodynamic (MHD) zoom-simulations carried out with a modified version of \ramses\ \citep{2002A&A...385..337T},
where we use sink particles as a sub-grid model of stars.
For a more detailed description of the setup and the procedure, please refer to our
previous works \citep[in the following referred to as K16 and K17]{2016ApJ...826...22K,2017ApJ...846....7K}.
A description of the modifications to \ramses\ and the implementation of our sink
particle model is also given in \citet{2012ApJ...759L..27P,2014ApJ...797...32P,2017arXiv170901078H} and \citet{2014IauS..299..131N}.
Our initial condition is a turbulent GMC modelled as a cubic box of (40 pc)$^3$ in size.
We use periodic boundary conditions.  
At an average number density in the box of 30 cm$^{-3}$, 
it contains approximately $10^5$ M$_{\rm \odot}$ of self-gravitating magnetized gas 
with an initial magnetic field strength of 3.5 $\mu$G.
The turbulence is driven self-consistently by massive stars exploding
as supernovae after a mass dependent lifetime
yielding a velocity dispersion of the cold 
and dense gas in agreement with Larson's velocity law \citep{1981MNRAS.194..809L}. 
In line with the recipe of \citet{1986PASP...98.1076F} for UV-induced heating \citep{2006agna.book.....O}, 
heating of the cold dense gas is quenched in the GMC. 
In less dense regions, we use a cooling function for optically thin gas based on the work by \citet{2012ApJS..202...13G}.
We evolve the GMC for about 5 Myr by using adaptive
mesh refinement with a root grid of $128^3$ and 9 AMR levels each refining the grid by a factor of 2.
Below we follow the usual convention in \ramses\ and quote the number with respect to the box size.
$128^3$ and 9 AMR levels corresponds then to having the global run evolved with a maximum refinement
of 16 levels, and a minimum cell size of 126 au.
During the GMC evolution, a few hundred stars
with different masses form throughout the GMC.
We refer to this run as the \emph{parental run}.
Afterwards, we select six sink particles that formed at different locations in the GMC, 
and evolved to about 1 to 2 solar masses by the end of the parental run at 5 Myr.
We use the zoom-simulation method and re-simulate smaller stretches in time of the parental run focusing
all computational resources on a single star -- while retaining the full (40 pc)$^3$ box -- to resolve the formation process and early
evolution. For the selected six cases, we use a maximum resolution of 22 levels of refinement,
corresponding to a minimum cell size of about 2 au for intervals of up to 200 kyr after sink formation.
Additionally, we re-simulate a sink with a massive disc approximately 50 kyr after sink formation
with high maximum resolution of 0.06 au, corresponding
to 27 levels of refinement ($0.06$ au) for 1000 yr.

\subsection{\mesa}
\label{sec:mesa}
Using as an input the time evolution of the accretion rates of the individual sinks in the zoom-simulations,
we model the evolution of the protostellar structure with the stellar evolution code \mesa{} \citep{2011ApJS..192....3P,2013ApJS..208....4P,2015ApJS..220...15P}.
The protostellar structure models are evolved using \mesa{} release 8845 and the code has been
augmented to support variable accretion rates and model the thermal efficiency at the accretion shock.
We refer to \citet{jensen}, where the same setup is used, for a detailed description.
The initial conditions of the model is a relaxed second Larson core of uniform composition.
Recent models of collapsing prestellar cores have found a universal second Larson core mass and radius of
$M \sim 0.0029$ M$_{\sun}$ and $R \sim 0.82$ R$_{\sun}$ for low- to intermediate-mass stars independent of the
pre-collapse conditions of the dense core \citep{2017A&A...598A.116V}.
Based on these results we opt for an initial core of mass $M = 0.0029$ M$_{\sun}$. 
However due to
differences in opacity tables we get a smaller initial radius of $R = 0.64$ R$_{\sun}$.
The metallicities of the second Larson core for hydrogen, deuterium
and helium are chosen to match the abundances found in the local interstellar
medium \citep{2001RvMP...73.1031F, 2000astro.ph..1409T,2010MNRAS.406.1108P} with mass fractions
$X=0.70$, $Y = 0.28$, $^{2}\mathrm{H} = 2\times10^{-5}$ and $^{3}\mathrm{He} = 2.98\times10^{-5}$.
For the remaining abundances we use the mass fractions from \citep{1998SSRv...85..161G}.
We accrete material with the same composition as the initial core.
If the protostar is fully convective material can be
carried to the centre, thereby continuously adding deuterium and lithium fuel to the protostar.
At low temperatures ($T < 4,000$ K) we use opacity tables from
\citep{2008ApJS..174..504F} while the tables of \citep{1998SSRv...85..161G} are used at higher temperatures.

Infalling material stalls at the accretion shock close to or at the stellar surface, where the potential energy is converted to heat.
This contributes both to the luminosity of the star and heats up the surface. It is an open question exactly how much energy is
radiated outwards, and how much is absorbed by the star. It depends non-trivially on the inflow geometry of the material,
the evolutionary stage of the star, and the level of accretion.
We adopt the same description of the accretion luminosity as \citep{2009ApJ...702L..27B}
\begin{equation}
L_{\rm acc}^{\rm in} = \epsilon \alpha \frac{G M \dot{M}}{R}, \quad L_{\rm acc}^{\rm out} = \epsilon (1-\alpha) \frac{G M \dot{M}}{R}\,,
\end{equation}
where $\alpha$ is the thermal efficiency parameter, $M$ and $R$ are the protostellar mass and radius respectively, $G$ is the
gravitational constant, $\dot{M}$ is the accretion rate, and $L_{\rm acc}^{\rm in,out}$ are the luminosities absorbed by the protostar and
radiated away from the protostar at the accretion shock respectively.
$\epsilon$ is dependent on the details of the accretion process, with $\epsilon \leqslant 1$ for gravitationally bound material and $\epsilon \leqslant 0.5$ for boundary layer accretion from a thin disc \citep{1997ApJ...475..770H}.
We set $\epsilon = 0.5$ as we assume disc accretion with half of the total accretion luminosity either lost to viscous heating in the disc or stored as
rotational energy.
For the radiative transfer simulations we assume that the luminosity of the accretion disc is exactly half of the accretion energy lost in the accretion disc:
\begin{equation}
L_{\rm acc}^{\rm disc} = \frac{G M \dot{M}}{4R}\,,
\label{eq:Lacc_disc}
\end{equation}
which is reasonable if accretion happens at the co-rotation radius \citep[see e.g.][]{jensen,1997ApJ...475..770H}

We use a simple qualitative model for the thermal efficiency parameter $\alpha$. It is a smooth step function that only depends on the instantaneous
accretion rate $\dot{M}$
\begin{equation}
\alpha(\dot{M}) = \frac{\alpha_{L} e^{\dot{M}_{m} / \Delta} + \alpha_{H} e^{\dot{M} / \Delta}}{e^{\dot{M}_{m} / \Delta} + e^{\dot{M} / \Delta}}\,,
\end{equation}
where $\alpha_{L} = 0.005$, $\alpha_{H} = 0.5$ are the lower and upper bounds of $\alpha$ and
$\dot{M}_{m}=3.5\times10^{-5}$ and $\Delta=(2.0/3.0)\times10^{-5}$ are the midpoint and the width of the crossover between $\alpha_{L}$ and $\alpha_{H}$.
Physically, it corresponds to assuming that in the high-accretion state the protostar is completely engulfed in the accretion flow, material is added
to the star at a high entropy, and the star is heated by the radiation from the accretion shock. In the low-accretion state material is added
in a process similar to magneto-spheric accretion, where it is funnelled in through accretion channels on to the star. At the foot point the stellar surface
is optically thick while the corona is optically thin. Excess heat is readily radiated away, and material is added to the star essentially at the same
entropy as the surface. See \citet{jensen} for a detailed discussion, and exploration of alternative models of the thermal efficiency.

The models include convection through the mixing length approximation with
$\alpha_{\rm MLT} = 1.82$ and convective core overshooting above and below convective zones.
For the convective core overshooting parameters we adopt the same values as \citep{2016ApJ...823..102C} for low- and intermediate-mass stars.

\subsection{\radmc}
\label{sec:radmc}

To produce synthetic observables of the simulated protostellar systems we use the dust radiative transfer code, \radmc\footnote{\url{http://www.ita.uni-heidelberg.de/~dullemond/software/radmc-3d/}} (version 0.41; see \citealt{Dullemond:2004iy} for a description of the two-dimensional version of the code). The calculation of synthetic observables requires knowledge about the dust temperatures. While the simulation use a tabulated equation of state, and includes heating from cosmic rays and a diffuse UV field, it does not include point-source radiative transfer. Dust temperatures are therefore computed with \radmc. \radmc uses the Monte Carlo method of \citet{Bjorkman:2001du} along with the continuous absorption method of \citet{Lucy:1999wd} to calculate the dust temperatures. The method computes the dust equilibrium temperatures by propagating a number of 'photon packets' through the model (set to five million in our case), and requires the inclusion of a dust opacity table as well as one or more luminosity sources.

For the dust opacities we use the monochromatic opacities of \citet{Semenov:2003hk}\footnote{\url{http://www2.mpia-hd.mpg.de/home/henning/Dust_opacities/Opacities/opacities.html}} corresponding to homogeneous spherical dust grains with normal iron content in the silicates ($\mathrm{Fe}/(\mathrm{Fe} + \mathrm{Mg}) = 0.3$). The \citeauthor{Semenov:2003hk} dust opacities are divided into five temperature ranges and we therefore calculate the dust temperatures over three iterations of the \citeauthor{Bjorkman:2001du} algorithm. In the first iteration we use the dust opacity table corresponding to the highest temperature range ($685\,\mathrm{K} \lesssim T_\mathrm{dust} \lesssim 1500\,\mathrm{K}$). In the two subsequent iterations the choice of dust opacity table is based on the cell temperature calculated in the previous iteration. For a typical model 99 \% of the cells change their temperatures by less than 10 \% between the second and third iteration and we therefore consider the temperatures to be sufficiently converged. The dust is assumed to sublimate at temperatures >\,1500\,K, which we achieve by setting the dust opacity to zero at all wavelengths for temperatures above this limit. This step prevents the dust temperatures from becoming artificially high in the innermost region of the model, which would have adverse effects both on the dust emission and on the calculated dust temperatures further out in the model. Finally, we assume a uniform dust-to-gas mass ratio of $0.01$ everywhere.

The \mesa models presented in Section~\ref{sec:mesa} calculate the protostellar luminosity and the accretion luminosity from the conversion of gravitational potential energy into heat at the accretion shock at the protostellar surface. The protostellar luminosity stems from the release of gravitational energy as the protostars contracts as well as energy released from the fusion of deuterium and lithium once the central temperatures reach the threshold values.
In the \radmc model these two luminosity components are added together and modelled as the emission from a perfect blackbody with an effective temperature as given by the \mesa models. Another source of emission is the viscous luminosity stemming from the angular momentum transport in the disc, however this emission is non-trivial to calculate self-consistently since the disc is not uniquely defined in the simulation and since the viscous luminosity itself also depends on the local dust temperature. We therefore use an estimate of the viscous luminosity from the \mesa models calculated using equation (\ref{eq:Lacc_disc}). Since the majority of the viscous luminosity originates from the innermost regions of the disc where both density and viscous stresses are highest \citep{Min:2011id} we simply model the viscous luminosity as a separate blackbody at the same position as the protostar. The effective temperature of the new blackbody is set to 1500\,K, which is the temperature of the dust just before it sublimates.

Synthetic spectral energy distributions (SEDs) are calculated adopting a square aperture with side length of 4000\,au, corresponding to roughly 10\,arcsec, at a distance of 420\,pc. For each \radmc model we calculate a set of 12 SEDs along different lines of sight to sample the full $4\pi$ solid angle as well as possible. To ensure an even distribution over the solid angle the lines of sight are arranged along the vertices of a icosahedron centred on the central source.

\section{Results}
\begin{table}
\centering
{
\begin{tabular}{c|c|c|c|c|c|c}

\begin{tabular}{@{}c@{}} Sink \end{tabular} & a & b & c & d & e & f \\ \hline
\begin{tabular}{@{}c@{}} \# in K17 \end{tabular} & 1 & 4 & 5 & 6 & 7 & 9 \\ \hline
\begin{tabular}{@{}c@{}} Disc? \end{tabular}  & \begin{tabular}{@{}c@{}} None/weak \end{tabular} & Massive & None & Massive & None & Weak \\
\end{tabular} }
\caption{Summary of the disc study in K17 for the six sinks discussed in this paper
showing whether a disc around the sink is strong, weak, or non-existing.}
\label{disc-overview}
\end{table}

\begin{figure*}
\subfigure{\includegraphics[width=0.98\linewidth]{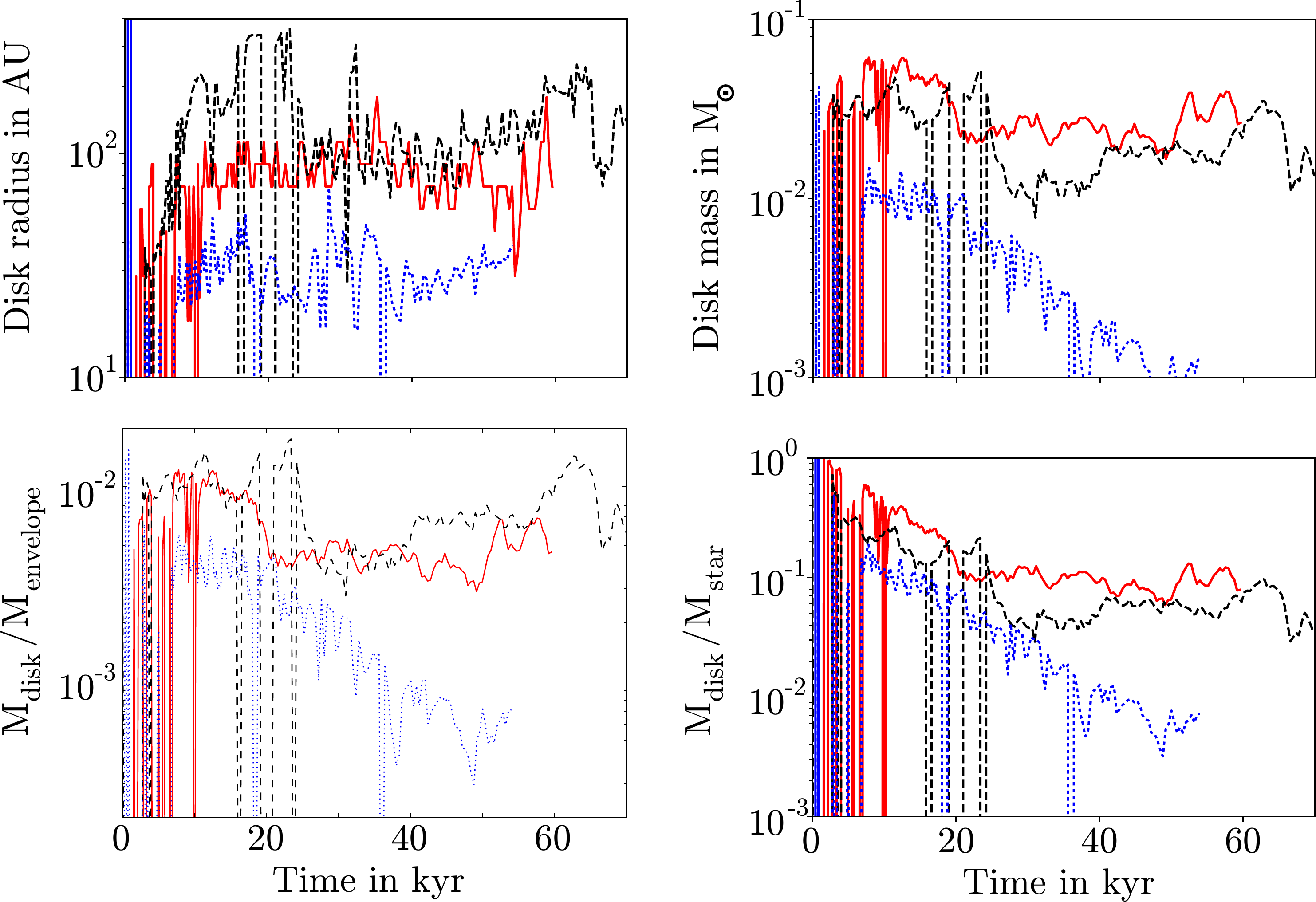} }
\protect\caption{\label{fig:disc_size_mass}
Disc radius (upper left panel), disc mass (upper right panel), ratio of disc and envelope mass $\frac{M_{\rm disc}}{M_{\rm env}}$ (lower left panel) and ratio of disc and stellar mass (lower right panel)
around sink b (red solid line), sink d (black dashed line)
and sink f (blue dotted line). 
The disc radius is estimated as the distance from the host sink,
where the rotational velocity is less than 80 \% of the Kepler speed $v_K = \sqrt{\frac{GM}{r}}$.
The disc mass is determined as the mass that is located within the disc radius and inside a vertical height from the mid-plane
computed as the maximum of 8 au and a vertical-height-to-radius ratio of 0.2.}
\end{figure*}

\subsection{Accretion profiles}
\begin{figure*}
\subfigure{\includegraphics[width=\linewidth]{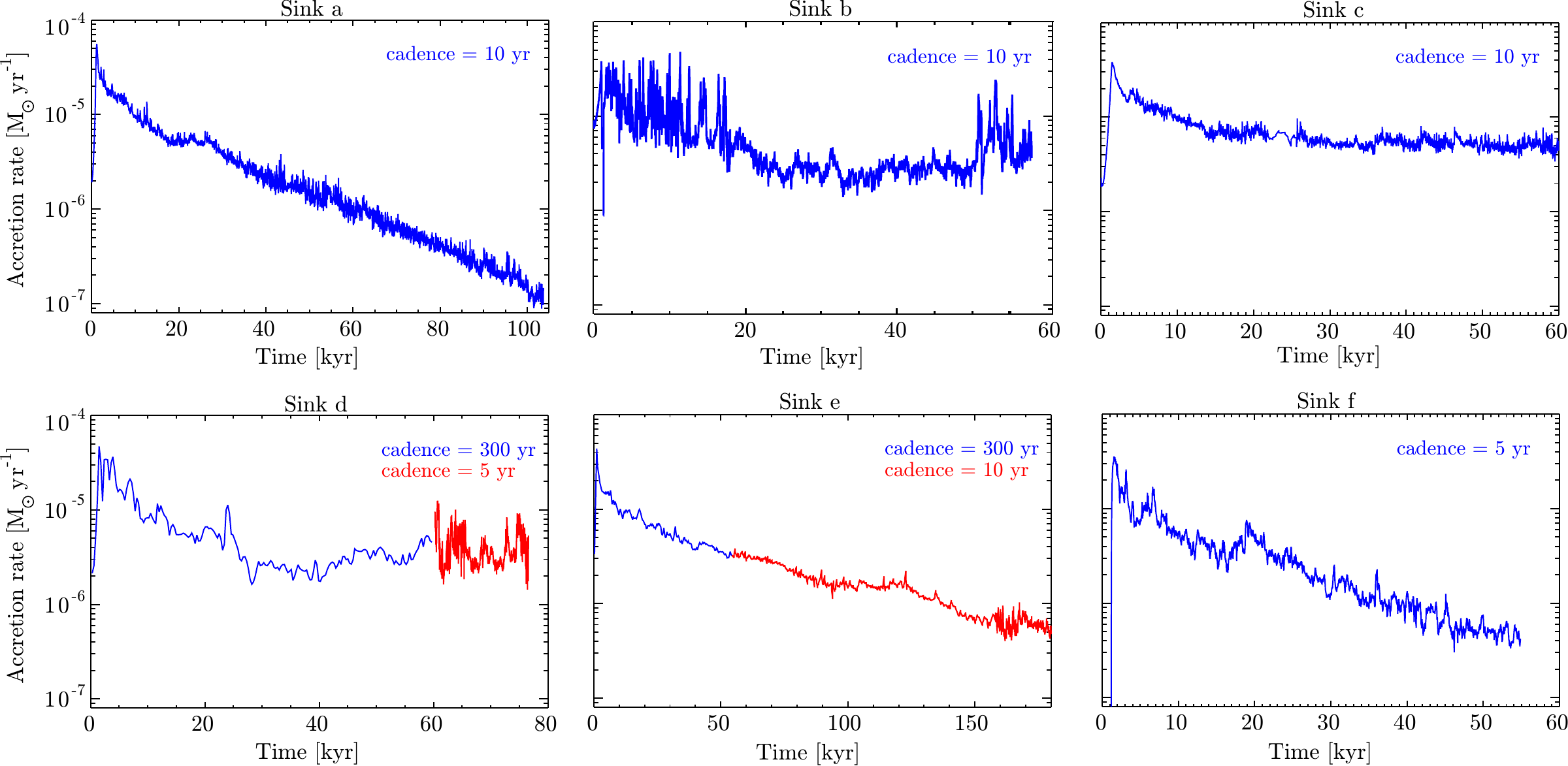} }
\protect\caption{\label{fig:acc-prof} Accretion rate in M$_{\sun}$yr$^{-1}$ for the different sinks as based on the runs carried out with a minimum cell size of 2 au.
The cadences for the individual sinks are given in the legend.
}
\end{figure*}
\begin{figure*}
\subfigure{\includegraphics[width=\linewidth]{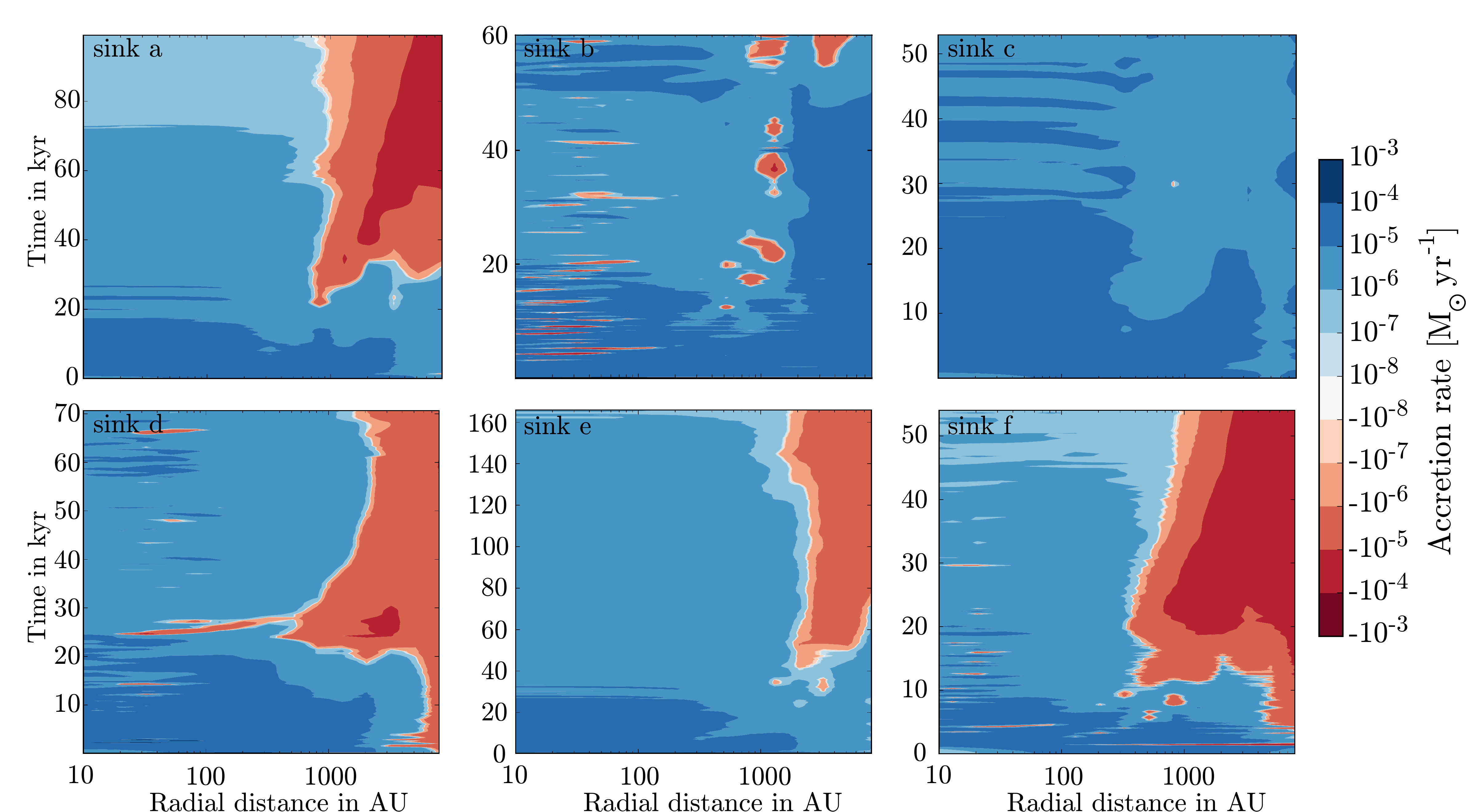} }
\protect\caption{\label{fig:acc-sphere} Mass change in solar masses per yr for the mass enclosed within spheres of different radii. The results are based on the runs carried out with a minimum cell size of 2 au.
}
\end{figure*}

In K17, we showed that the accretion profiles of different protostars differ
from each other due to the molecular cloud environment of the protostellar systems.
In this paper, we investigate the effect of an existing protoplanetary disc in the vicinity of the sinks in more detail.
To compare the discs around the different sinks,
we defined the radius of the disc in K17 as being equivalent to the radial distance from the sink,
where the rotational speed dropped to less than 80 \% of the Kepler-speed $v_K = \sqrt{GM/r}$.
We excluded the inner seven cells (14 au) from the sink 
to avoid accounting for the low rotational velocities induced by the sink at the centre. 
We then estimated the disc mass to be the enclosed mass within this radius and inside a vertical height from the mid-plane
computed as the maximum of 8 au and a vertical-height-to-radius ratio of 0.2.
Although these criteria only give a rough estimate of the disc radius and mass, we found clear differences for the different sinks.
Around two sinks (sink b and d), discs of several tens of au are formed
within a few kyr and evolve to large disc-to-stellar mass ratios $\left( \frac{M_{\rm disc}}{M_{\rm star}} \right)$
of about 10 \% at 50 kyr after sink formation.
Another sink (sink f) also showed signs of a stable, but smaller and lighter disc (only a few tens of au and $\frac{M_{\rm disc}}{M_{\rm star}}$ of only 1 \% at about 50 kyr after sink formation),
whereas the vicinity of the remaining sinks either showed no or only weak and intermittent signs of a disc.
In Table \ref{disc-overview}, we briefly summarize the main result of the disc study carried out in K17
by distinguishing between no disc, a weak disc or a massive disc, and in the upper panels of \Fig{disc_size_mass}
we show the disc radius (left) and disc mass (right) around sink b, d and f.
Moreover, we show in the lower panels of \Fig{disc_size_mass} the ratio between disc mass and envelope mass (left) as well as the ratio between disc mass and sink mass (right). 
In principle the envelope mass is all the mass that is gravitationally bound inside the collapsing core.
As shown in K17, the pre-stellar cores in our simulations are not perfectly symmetric spheres
due to the underlying turbulence and the cores are shaped differently. 
However, as a proxy we simply define the envelope mass as all the mass that is located within 8000 au from the central sink
except for the sink mass itself. 
We find that the three sinks have envelope masses of a few to about 10 M$_{\rm \odot}$. 
Comparing the evolution of the disc mass with the evolution of $M_{\rm disc}/M_{\rm envelope}$
does not indicate a simple correlation between envelope and disc mass.   
The disc mass around sink b tends to be larger than the disc mass around sink d, 
but the ratio of $M_{\rm disc}/M_{\rm envelope}$ is generally lower during the evolution.
Hence, the plot reveals that sink b has the most massive environment.

\subsection{Accretion profiles of the six sinks}
In \Fig{acc-prof} we show the accretion profiles of the six sinks in our zoom-simulations.
Except for the first 60 kyr around sink d and approximately the first 55 kyr around sink e,
we trace the accretion of mass on to the sink with higher cadence in time
(with 5 or 10 yr instead of 200 or 400 yr as in K17),
and thus, we are able to resolve accretion bursts of shorter duration.
The cases with the most massive discs -- sink b and sink d (sink 4 and sink 6 in our previous study) --
show the strongest fluctuations (of more than a factor of 5) in the accretion profile compared to sinks without discs
or with only weak discs.
The accretion profiles of sink b show fluctuations in the accretion profile of more than a factor of 10 
within time intervals of less than 100 yr.
In particular, sink b accretes mass intermittently.
The sink shows violent phases with strong fluctuations in the accretion rate, and calmer periods with modest fluctuations,
both lasting for several to tens of kyr. 
In comparison to previous studies explicitly studying episodic accretion in more idealized setups without magnetic fields \citep[e.g][]{2010ApJ...723.1294V,2011ApJ...729..146V,2011ApJ...730...32S}
, the occurrence of episodic events is lower in our cases. 
Considering that magnetic fields limit the growth of disks due to magnetic braking, 
as well as they provide additional pressure support against fragmentation these differences are expected.

In addition to the accretion profile on to the central sink,
we are interested in the mass evolution at different distances from the central star.
We computed the change in total mass enclosed in spheres (including all the mass that has fallen on to the central sink)
of radius $r$ during time intervals of 200 or 400 yr.
In \Fig{acc-sphere} we show the corresponding result in the range of 10--8000 au.
As expected, the mass accretion profiles for the innermost radii are similar to the accretion profiles of the sinks.
For the cases without discs (sinks a, c, and e), we continuously see mass increase
throughout the entire period of the simulation within distances of about 1000 au.
However, for the sinks with surrounding discs (sinks b, d, and f),
we see brief periods in which the mass enclosed in spheres of a few tens to about several 100 au occasionally decreases.
The intermittent decrease of enclosed gas at these scales is caused
by temporary outward motion of the gas.
Finally, we find that the mass accretion on scales beyond 1000 au differs significantly in the six systems,
and is fundamentally shaped by the molecular cloud environment.
After an initial increase in total enclosed mass during the first several kyr after sink formation,
the total enclosed mass around sinks a, d, e and f decreases with a rate of up to $10^{-4}$ M$_{\sun}$yr$^{-1}$ for spheres
with radii larger than $\approx$1000 au.
The early total mass increase within small radii traces
the collapse of the prestellar core and the subsequent infall of gas from the envelope on to the star--disc system
during the first 10 to 20 kyr after the birth of the protostar.
In contrast, the total mass evolution within spheres with radii larger than the size of the prestellar core
can only increase if additional mass is delivered.
The change in mass for spheres with radii larger than $\sim$1000 au
trace the motion of the star--disc system through the GMC to regions of different densities.
In the cases of sink a, d, e and f, the star--disc system moves to a lower density region,
as indicated by the mass decrease at large radii, whereas for sinks b and c,
the systems are embedded in regions of higher density.
We caution that although the overall mass enclosed within a sphere of several 1000 au in radius may decrease
temporarily, the protostellar system may still be fed with additional mass from larger distances.

\begin{figure}
\subfigure{\includegraphics[width=\linewidth]{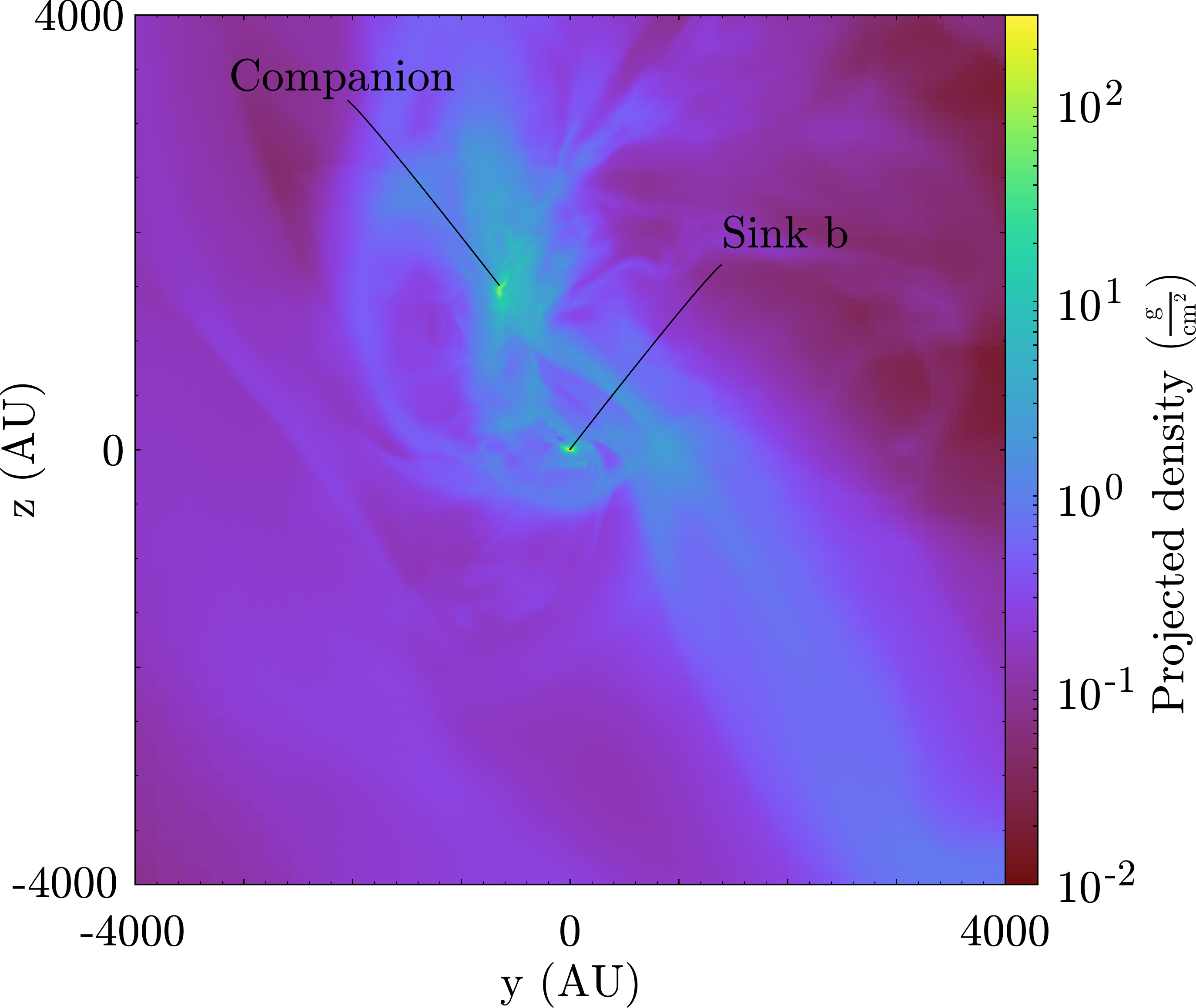} }
\protect\caption{\label{fig:wide_binary} The formation of a wide binary.
Column density in the $yz$-plane of the coordinate system with a column height
in the $x$-direction of $\pm8000$ au at $t = 35$ kyr after the formation of sink b.
Sink b is located at the centre and the companion forms
inside the massive clump locate at a relative position of $(x,y,z)\approx (-350, -600, 1250)$ au.
}
\end{figure}

In K17 we showed that the stars are located in different environments.
The system around sink b is embedded in a turbulent environment
and 3D visualizations with the tool \vapor\ \citep{clyne2005prototype,clyne2007interactive} show that additional mass approaches the star--disc system.
Moreover, a massive clump forms at a distance of about 1500 au from sink b that leads to the formation of
a companion about 35 kyr after the formation of sink b.
The formation of the companion reflects that sink b is embedded in a dense environment
as shown in the column density plot for sink b in \Fig{wide_binary}.
While it is very interesting to catch the formation and evolution of a wide binary system from realistic GMC conditions in a model
with high enough resolution to capture au-scale structures like the circumstellar disc, a detailed analysis is beyond the scope of this paper.
We will analyse the properties and dynamics of this system in detail in a follow-up paper.

\begin{figure}
\subfigure{\includegraphics[width=\linewidth]{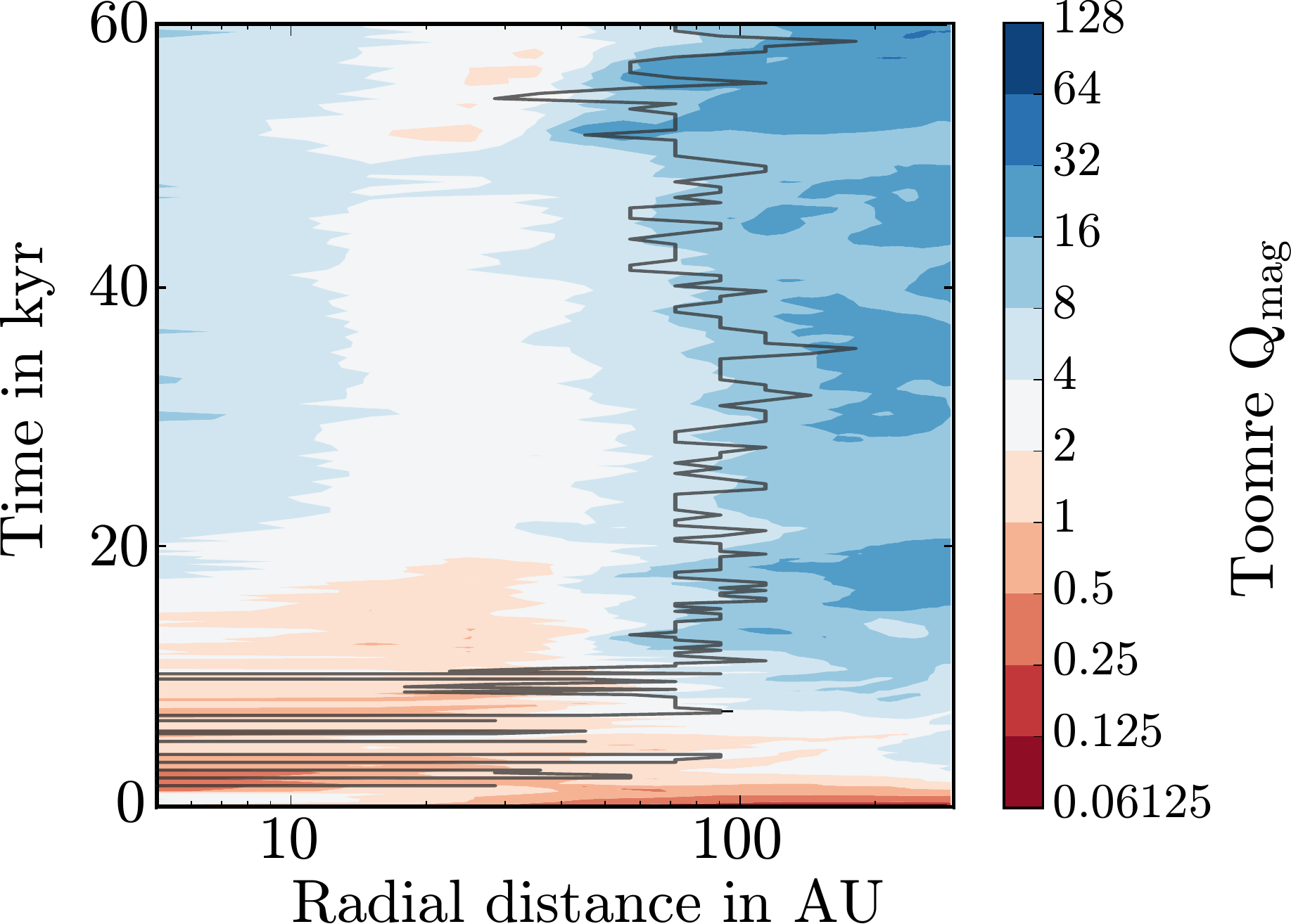} }
\protect\caption{\label{fig:toomre_contour} Azimuthally averaged magnetic Toomre parameter $Q_{\rm mag}$
around sink b during disc evolution.
The colour bar indicates the strength of $Q_{\rm mag}$ and
the solid grey line marks the size of the disc estimated as the radius, where the rotational speed drops to less than
80 \% of the Kepler speed $v_K = \sqrt{\frac{GM}{R}}$.
}
\end{figure}

\begin{figure*}
\subfigure{\includegraphics[width=\linewidth]{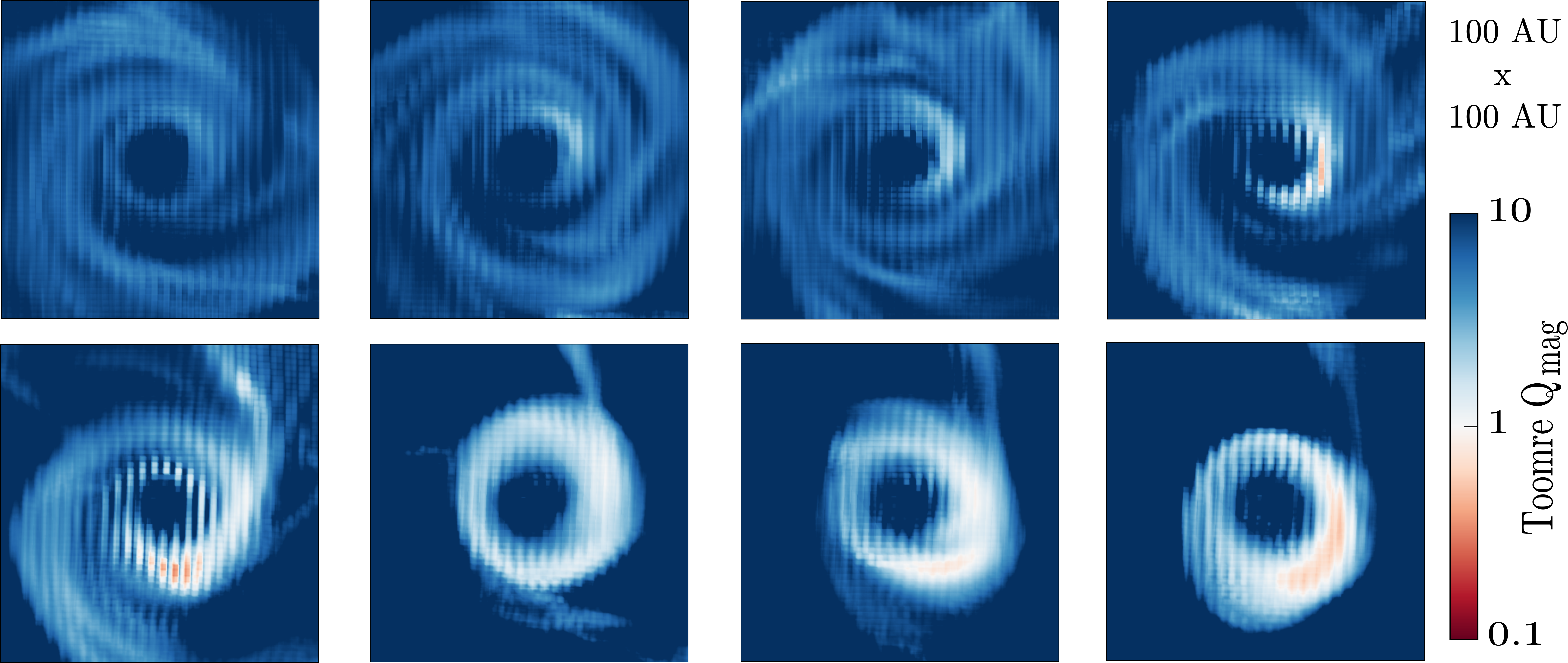} }
\protect\caption{\label{fig:toomre_proj} A $100\unit{au}\times 100\unit{au}$ image of the magnetic Toomre parameter $Q_{\rm mag}$
around sink b in the disc at $t=49.6$ kyr, $t=50$ kyr, $50.4$ kyr, $t=50.8$ kyr,
$51.2$ kyr, $51.6$ kyr, $52$ kyr and $52.4$ kyr.
The colour bar indicates the strength of $Q_{\rm mag}$.
}
\end{figure*}

\subsection{Disc instabilities}
\label{sec:discinstabilities}
\subsubsection{Low resolution run with minimum cell size of 2 au}
To better understand the influence of the disc on the accretion process, we investigate the stability of the disc around sink b in more detail.
Generally, if the ratio of the disc to stellar mass is high enough, it can exceed the support provided by thermal and magnetic pressure, and thus, the disc may become gravitationally unstable.
The stability parameter for a magnetized disc is the magnetic Toomre parameter \citep{1964ApJ...139.1217T,2001ApJ...559...70K}
\begin{equation}
Q_{\rm mag} = \frac{\sqrt{(c_s^2 + v_A^2)} \Omega}{\pi G\Sigma},
\end{equation}
with sound speed $c_s$, Alfv\'{e}n speed $v_A$, Keplerian orbital angular velocity $\Omega$,
gravitational constant $G$ and surface density $\Sigma$.
The disc is gravitationally unstable if $Q_{\rm mag} \lesssim 1$.
For detailed analyses of the critical number relevant for collapse in non-magnetized local disc simulations
addressing the
gravitational instability, please refer to \citet{1999ApJ...521..650B,2011ApJ...740....1K,2015ApJ...814..155B}.

\begin{figure}
\subfigure{\includegraphics[width=\linewidth]{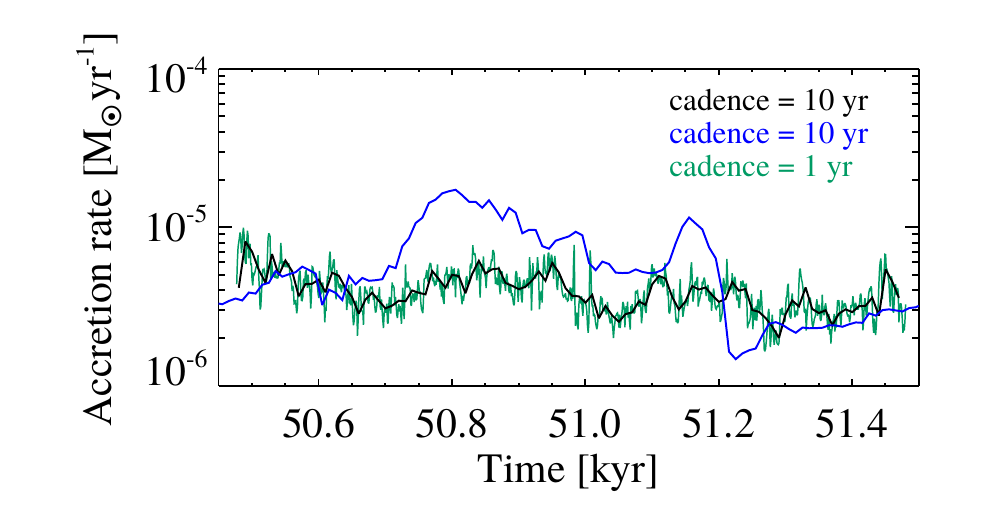} }
\protect\caption{\label{fig:acc27_prof} Accretion rate for the high-resolution run (black line)
and the low resolution run (blue) around sink b based on the same cadence of 10 yr for comparison.
The thin green line shows the accretion rate for the high-resolution run at the higher cadence of the
underlying data of 1 yr.
}
\end{figure}

\begin{figure}
\subfigure{\includegraphics[width=\linewidth]{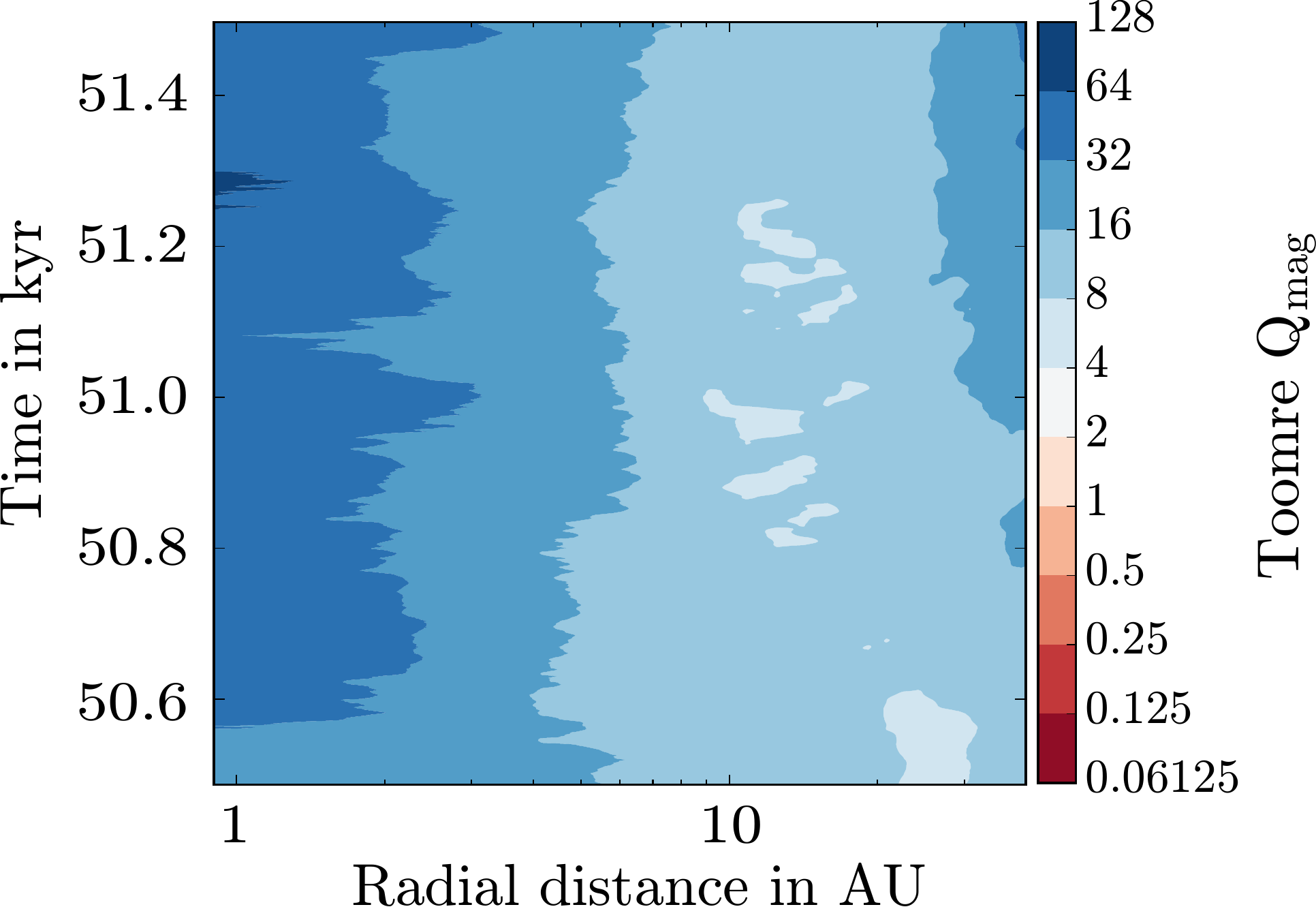} }
\protect\caption{\label{fig:toomre27_contour} Azimuthally averaged magnetic Toomre parameter $Q_{\rm mag}$
around sink b for the high-resolution run with minimum cell size of 0.06 au
for a period of 1 kyr around 51 kyr after sink formation.
The colour bar indicates the strength of $Q_{\rm mag}$.
}
\end{figure}

\begin{figure*}
\subfigure{\includegraphics[width=\linewidth]{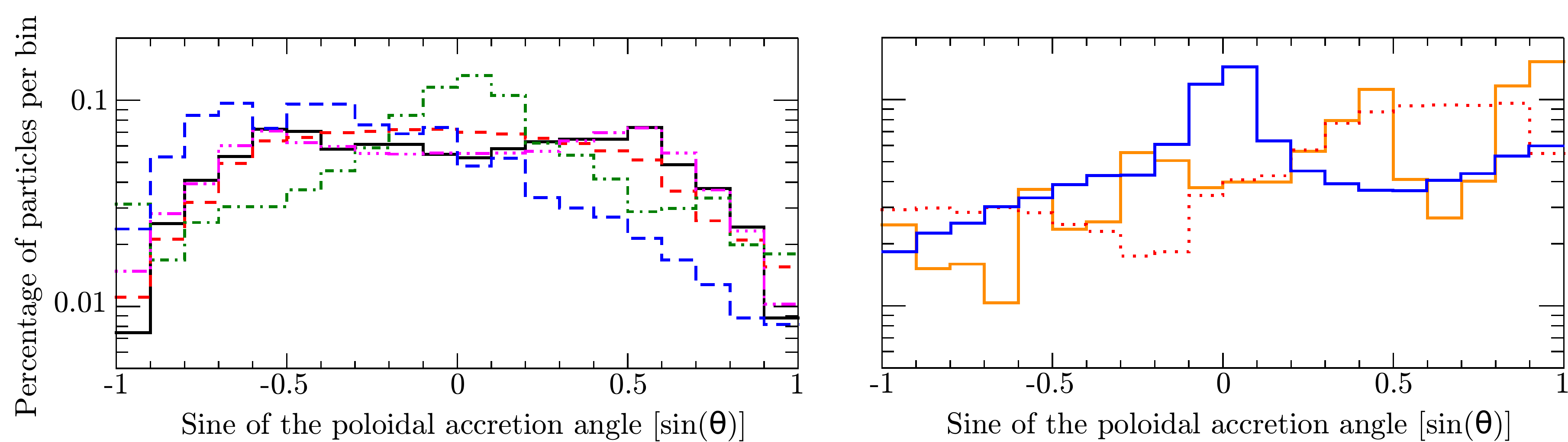} }
\protect\caption{\label{fig:part_acc_angle_distribution}
Left-hand panel: Probability distribution of the accretion angle for the different sinks
a: black solid line, c: red dashed line, d: green dashed-dotted line,
e: magenta dashed-double-dotted line, f: blue long-dashed line).
Right-hand panel: Probability distribution for sink b (blue solid line)
and the 1 kyr interval shortly after 50 kyr in the
2 au resolution run (orange solid line) and the high-resolution run (red dotted line).
}
\end{figure*}

In \Fig{toomre_contour}, we show a space--time diagram of the azimuthally averaged magnetic Toomre parameter
based on cylindrical shells of $100$ au in height. The figure shows azimuthal averages that indicate occasionally gravitationally instabilities at some locations in the disc.
Comparing \Fig{toomre_contour} with the accretion profile of sink b (upper middle panel in \Fig{acc-prof},
we find that the occurrence of low $Q_{\rm mag}$ (i.e.\ potential gravitational instabilities) correlate with periods
of high-accretion rates. This is expected, since angular momentum transport in an unstable disc is efficient.
During the first $\sim20$ kyr, the system is generally unstable or marginally stable with $Q_{\rm mag} < 2$, and the
accretion rate shows a fluctuating profile with strong bursts. 
These results are consistent with the evolution of the Toomre parameter around sink d as discussed in K17.
While indicative, the Toomre parameter is ill-defined, when
no disc is present, and low $Q_{\rm mag}$-values in the early collapse phase should not be over-interpreted.
From about 20 kyr to 50 kyr, $Q_{\rm mag}$ is above 2,
and the accretion profile is comparatively steady during this period.
Between $t=50$ and $60$ kyr, $Q_{\rm mag}$ decreases below 2 at a distance of 20 au,
and the sink undergoes a phase of more violent accretion for several thousand yr.

To investigate the spatial distribution of $Q_{\rm mag}$ in more detail,
we show the distribution in slices of 100 au $\times$ 100 au of the disc
for eight snapshots with 400 yr difference in between $t=49.6$ and $52.4$ kyr in \Fig{toomre_proj}.
As anticipated from the azimuthally averaged $Q_{\rm mag}$ profile and from the accretion profile,
we find that $Q_{\rm mag}$ locally becomes less than 1 at distances of 10 -- 20 au from the sink after $50.8$ kyr.
The rapid short-term fluctuations detected in the accretion profiles
indeed correlate with the occurrence of gravitational instabilities in the disc as suggested by \citet{2011ApJ...730...32S,2012MNRAS.427.1182S,2012ApJ...747...52D,2015ApJ...805..115V}.
However, in our study the instabilities trigger spiral structures in the disc, but fragmentation does not occur. 
This may be because the
areas where $Q_{\rm mag}<1$ are very localized.
The mass accretion rate within a spherical distance of several 100 au  increases from
$10^{-6}$ to $10^{-5}$ M$_{\sun}$ yr$^{-1}$ briefly before and during the burst period,
and the disc instability is therefore triggered by infall on to the star--disc system.
This scenario is different from the occurrence of an instability in a smooth disc of constant mass.

\begin{figure*}
\subfigure{\includegraphics[width=0.99\linewidth]{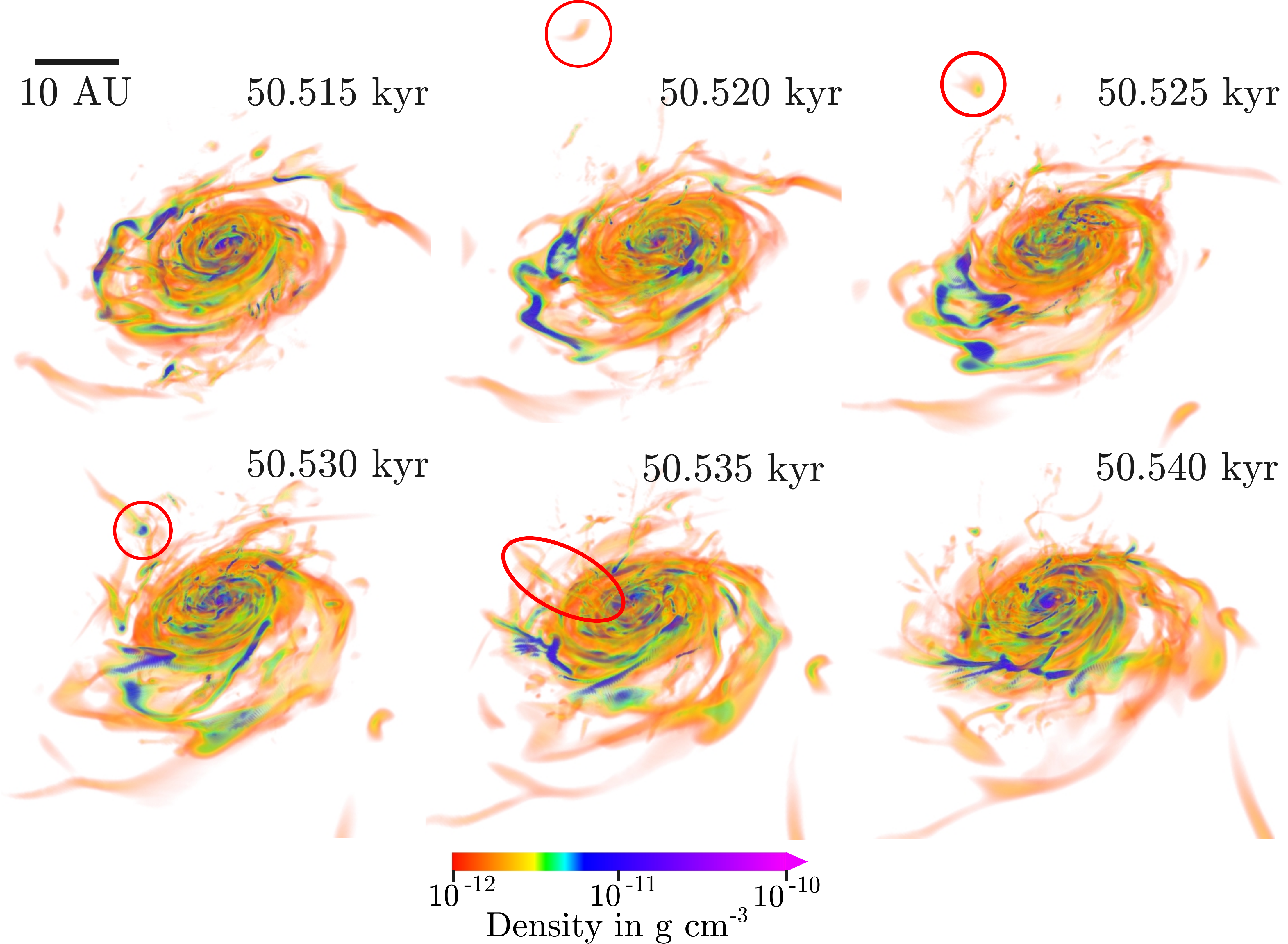} }
\protect\caption{\label{fig:disc_high_res}
Illustration of gas falling vertically on to the disc at a radius of $\sim 1$ au at $t= 50.515$\,kyr
based on the high-resolution run with a minimum cell size of $0.06$\,au.
The image was rendered with \vapor\ and the visible disc is about 10 au in radius. 
We chose a non-linear density dependent opacity such that low densities are more or even fully transparent,
and the disc structure is as apparent as possible. 
Therefore, the outer parts of the disc are not visible.
The red circles/ellipses illustrate the location of the blob at different times and 
the time cadence between the images is 5 yr.}
\end{figure*}

\subsubsection{high-resolution run with minimum cell size of 0.06 au}
To gain further insight into the accretion process and to test the robustness of the instabilities in the disc
around sink b,
we reran a sequence around 51 kyr with higher resolution (minimum cell size 0.06 au, 27 levels of refinement) for 1 kyr.
Similar to the zoom-in process before sink formation,
we gradually increased refinement level-by-level to continuously allow the cells to adopt to the higher resolution,
hence the discrepancy of the accretion profile shortly after $50.5$ kyr.
The overall accretion rates are similar to the ones in the lower resolution run (\Fig{acc27_prof}).
However, the accretion profile is more flat than in the lower resolution case.
This is because the disc is stable against gravitational instabilities
with generally higher azimuthally averaged magnetic Toomre parameters of at least $Q_{\rm mag}=4$
during the high zoom-in run as shown in \Fig{toomre27_contour}.
Therefore, the accretion bursts induced by the gravitational instabilities in the 2 au resolution run 
are absent in the run with minimum cell size of $0.06$ au.
In fact, the higher resolution run shows a smooth, radially decreasing density profile,
while in the lower resolution run the infalling mass cannot be transported properly to the inner disc
due to the lacking resolution.
As a consequence, mass piles up in the disc just outside the accretion region of a few cells (i.e. $\sim$10 au)
and builds an artificial mass buffer, which eventually causes gravitational instabilities in the disc.
We point out that the cumulative mass at $r\gtrsim 20$ au agree very well in both the low and high-resolution run
showing the overall consistency in our runs.

\subsection{Angular distribution of the accretion flow}
When matter is accreted from the disc or the envelope on to the star potential energy is transferred into kinetic energy,
and the kinetic energy is
either released as heat at the accretion shock or stored in the rotational energy of the star. This can happen through magneto-spheric accretion
channels, which funnels matter from the disc to the star, or in a boundary layer on top of the protostar. How much of the associated accretion
luminosity is absorbed by the star and how much is radiated
away remains an open question and probably depends on the conditions of the shock front and the geometry of the accretion flow.
In the case of a spherical and opaque accretion flow, the star would heat up because it cannot re-emit the radiation in any direction.
However, stars may accrete their mass through accretion channels covering only a fraction of the surface.
If the region outside the accretion channels is optically thin, the star can efficiently irradiate its accretion energy in the non-accreting directions.
We already showed in K17 and in agreement with \citet{2013MNRAS.432.3320S,2015MNRAS.446.2776S} that
mass accretion is filamentary -- except for the early collapse phase lasting for a few kyr.

We further constrain the direction from which the mass accretes on to the star.
In \Fig{part_acc_angle_distribution}, we illustrate the probability distribution of the accretion angles
as based on the location of tracer particles before they accrete on to the six sinks.
On top of that we show the angle distribution obtained from the higher resolution run for sink b
and the corresponding distribution for that interval as based on the standard resolution run.
We find that the sinks with the strongest discs (sinks b and d) predominantly accrete mass
at a small or zero angle with respect to the disc (enhancement at  $\sin(\theta)=0$),
showing that mass indeed accretes on to the star preferentially from the disc.
Generally, we can see for the remaining sinks that mass accretes less likely from the poles
as seen in the decrease of the profiles around $\sin(\theta)=\pm 1$.
In the cases of sinks a, c, d, and e the angle distribution is approximately symmetric,
while for sink b and particularly for sink f the distribution shows that gas predominantly accretes
from one side rather than the other.
As mentioned above,
gas accretion is mostly filamentary and occurs through accretion channels. Using 3D visualization
we have validated this, and that gas falls predominantly on to the central region in a plane that is not aligned
with the star--disc system during this period, explaining the asymmetry in the angle distribution.
When looking at the short interval that was run with higher resolution around sink b,
the asymmetry found in the low resolution run persists.
We show in \Fig{disc_high_res} the vertical infall of gas on to the inner part of a young disc at
$t\approx 50 $ kyr based on the high-resolution run.
Gas approaches the disc, forms a gas blob and accretes on to the disc at a radius of $\sim 1$ au.
Similar events occur preferentially from one side as a consequence of mass infall
explaining the asymmetry in the polar angle during the interval.
Besides the asymmetric infall, \Fig{disc_high_res} illustrates the underlying density structure inside the disc
and its evolution during a few to several orbital times of the inner disc.
The images demonstrate the importance of accounting for the environment by using global models when studying young discs.

\subsection{Effect of episodic accretion on protostellar evolution}
Accretion bursts can have a profound impact on the protostellar evolution. When the rate of mass accretion
suddenly increases, a much larger fraction -- maybe all -- of the stellar surface is covered with accreting material \citep{1996ApJ...461..933K}.
This inhibits the effective radiation and cooling of material before it is added to the star.
A significant influx of heat together with new material disrupts the local equilibria that dominate the stellar dynamics
\citep{2016A&A...588A..85G,jensen}.
In this section we describe the effects of episodic accretion on the stellar evolution.

\begin{figure*}
\subfigure{\includegraphics[width=\linewidth]{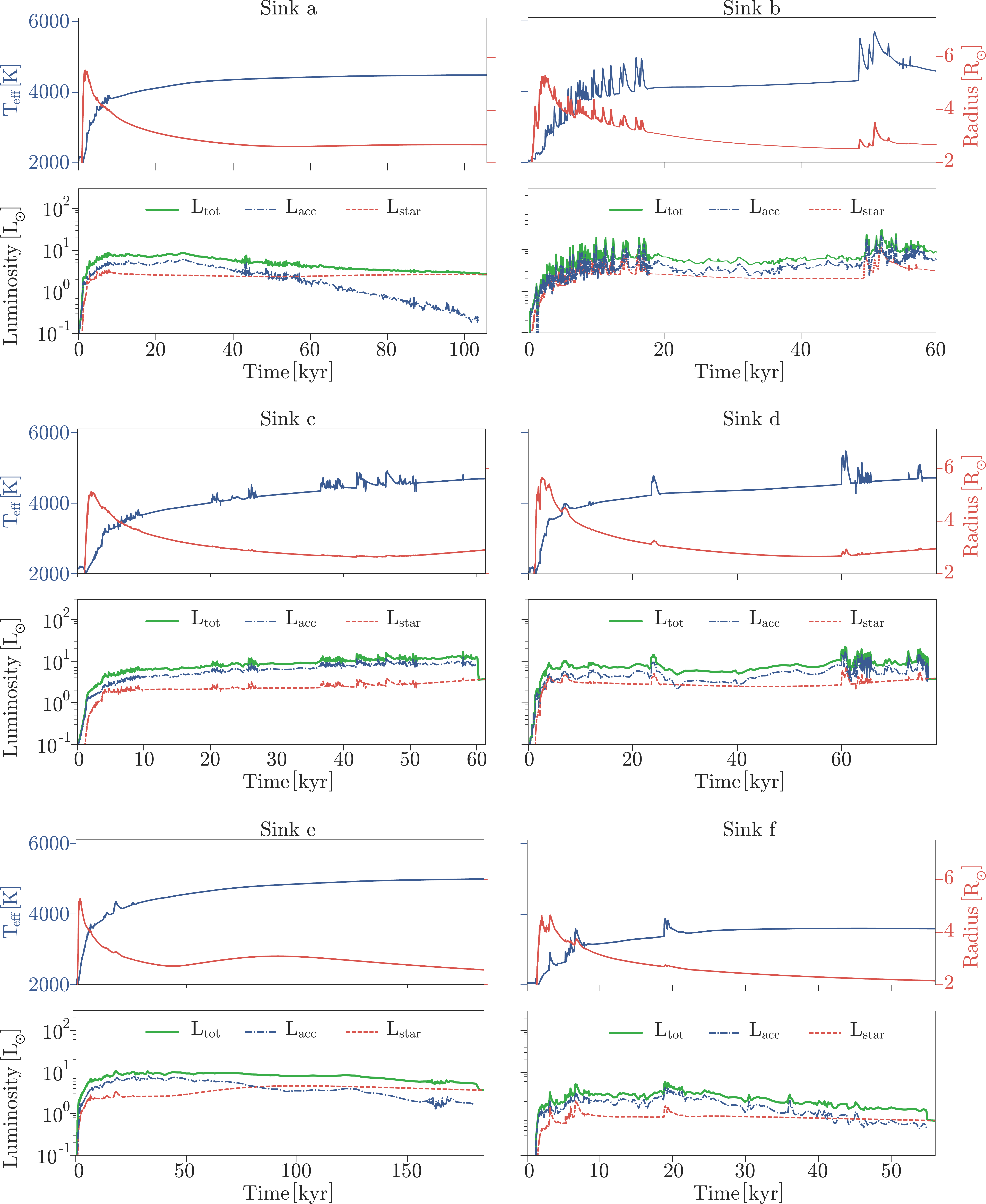} }
\protect\caption{\label{fig:mesa_figures} Stellar properties computed with \mesa based on the accretion profiles of the six sinks shown in \Fig{acc-prof}.
Panels in rows 1, 3, and 5: Evolution of effective temperature (blue colour) and radius (red colour) of the protostar;
Panels in rows 2, 4, and 6: Evolution of the stellar (red dashed), accretion (blue dashed--dotted) and total (green solid) luminosity.
}
\end{figure*}

\Figure{mesa_figures} shows the evolution of the protostellar radius, effective temperature and various luminosity
components during the accretion phase.
We start by focusing on the evolution of sink a, sink e, and sink f, 
which do not feature accretion events with $\dot{M} > 10^{-5}\,\Msunyr$
of any significance at later times.
During the first 10 kyr the protostars undergo a rapid expansion phase caused by the deposition of accretion energy in the outer layers.
Due to the dynamical $\alpha(\dot{\mathrm{M}})$ model these high initial accretion rates lead to a high thermal efficiency with $\alpha \approx 0.5$.
In combination with a small initial radius and high-accretion rates this gives a relatively high $L_{\mathrm{acc}}^{\mathrm{in}}$ compared to protostellar luminosity $L_{*}$.
During this period the thermal energy of the accreted material heats the surface layers of the protostars faster than the energy can be radiated away, which causes a readjustment of the hydrostatic structure.
For a comprehensive description of the conditions behind this mechanism we refer to \citet{1997ApJ...475..770H}.
This self-regulating mechanism of the initial protostellar structure for hot accretion models removes the dependence on the initial radius of the model, but introduces a new dependence, namely the efficiency parameter $\alpha$ and the accretion rate \citep{2011ApJ...738..140H}.
The expansion is halted as the increased radius of the protostars causes the protostellar luminosity to rise and counterbalance the heat influx
from the accreted gas. As the main accretion phase ends the protostar begins an almost steady contraction. At the beginning the stars are still
not in a stable hydrostatic configuration, and keep increasing the effective temperature, while eventually around $\sim$100 kyr they start to
descend the Hayashi track and evolve similar to that of a classical pre-main-sequence model. It is also around this time the protostars ignite deuterium in the interior.

Next we turn our attention to the evolution of sink b, which is dominated by a number of intense accretion burst.
The initial behaviour of the model is similar to that of sink a with a rapid readjustment of the hydrostatic equilibrium.
After this initial phase we see a number of accretion events with $\dot{M} > 10^{-5} \,\Msunyr$.
Each of them are followed by an increase in radius and effective temperature,
which is a consequence of the same mechanism responsible for the initial expansion.
Due to the dynamical $\alpha$-model the accretion is hot with $\alpha$ approaching $0.5$ during an accretion event.
The hot accretion combined with the high-accretion rates causes a significant rise in $L_{\mathrm{acc}}^{\mathrm{in}}$, which heats the outer
layers of the protostar and prompts a readjustment of the stellar structure.
This behaviour during accretion events with hot accretion is in agreement with the results of \citet{2012ApJ...756..118B},
who also studied the effects of accretion burst on accreting protostars.
After the accretion event the protostar quickly contracts and settles down to the previous configuration.
The accretion event pauses the contraction of the protostar, and delays the approach to the main sequence, however the effective delay is small
compared to the overall timescale of star formation.
The effects of episodic accretion on the timescales of star formation is analysed in detail by \citet{jensen}.

The stellar expansion and rise in effective temperature leads to a higher stellar luminosity,
which, combined with the rise in outward accretion luminosity during such a burst, can lead to significant
changes in the total luminosity of the protostar.

\subsection{Observational signatures of episodic accretion in the dust continuum}

An accretion burst is expected to manifest itself by increased continuum emission at all wavelengths with a response time of days or hours at infrared wavelengths, and a response time of up to a few months at sub-mm wavelengths \citep{Johnstone:2013je}. Given that classical FUor outbursts are generally expected to have durations >10\,yr the timescale is short enough that it will be possible to directly detect the increased luminosity due to an accretion burst, even in deeply embedded objects that can only be detected at long wavelengths.

\begin{figure}
\subfigure{\includegraphics[width=\linewidth]{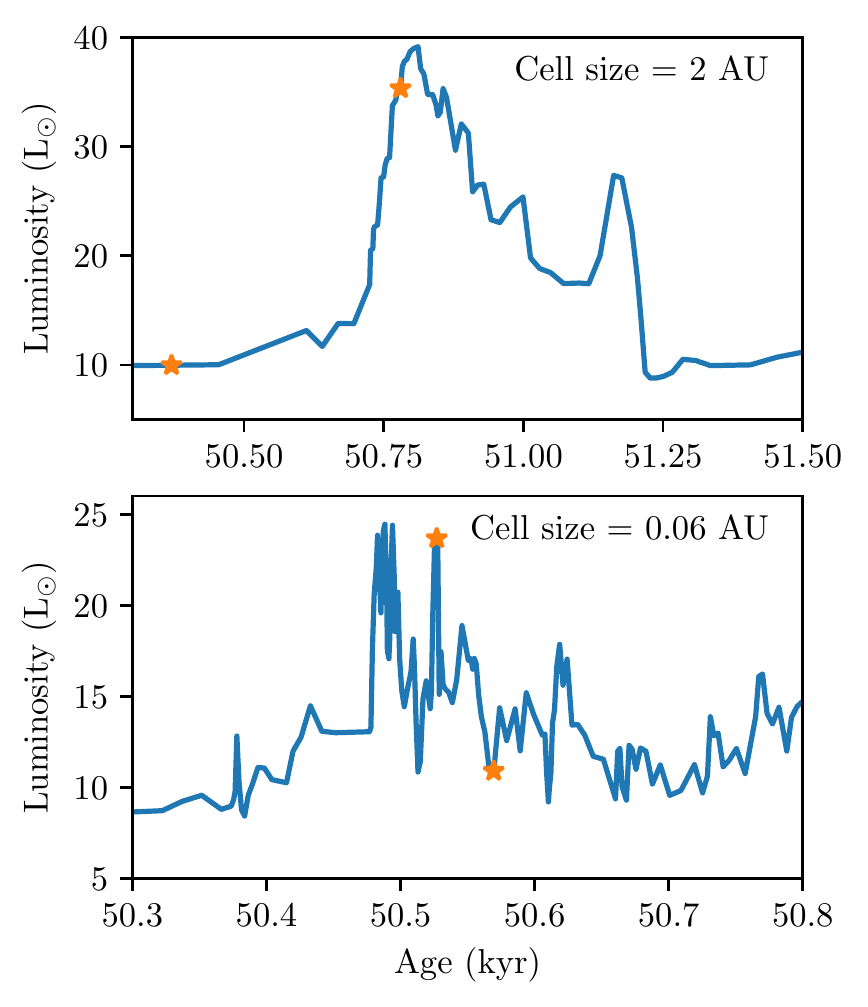} }
\protect\caption{\label{fig:luminosity} Total luminosity (accretion plus intrinsic) of sink~b for both the 2\,au resolution case (top) and 0.06\,au resolution (bottom). The stars on the curves represent the snapshots where the radiative transfer models were calculated.}
\end{figure}

Here, we study the variation in the continuum emission in the sink~b system, which shows the most accretion variability. We focus on the 1\,kyr stretch of the simulation that was rerun with 27 levels of refinement corresponding to a minimum cell size of 0.06\,au. As discussed in section~\ref{sec:discinstabilities} the accretion profiles differ between the low- and high-resolution runs because of more efficient angular momentum transport in the high-resolution disc. \Fig{luminosity} shows the total luminosity (intrinsic luminosity plus accretion luminosity) of sink~b for the two available resolutions. The 2\,au resolution run shows a regular long-lived accretion burst, while the 0.06\,au resolution run also shows a high degree of variability, but the bursts are weaker and have shorter duration relative to the 2\,au resolution case.

\begin{figure*}
\subfigure{\includegraphics[width=\linewidth]{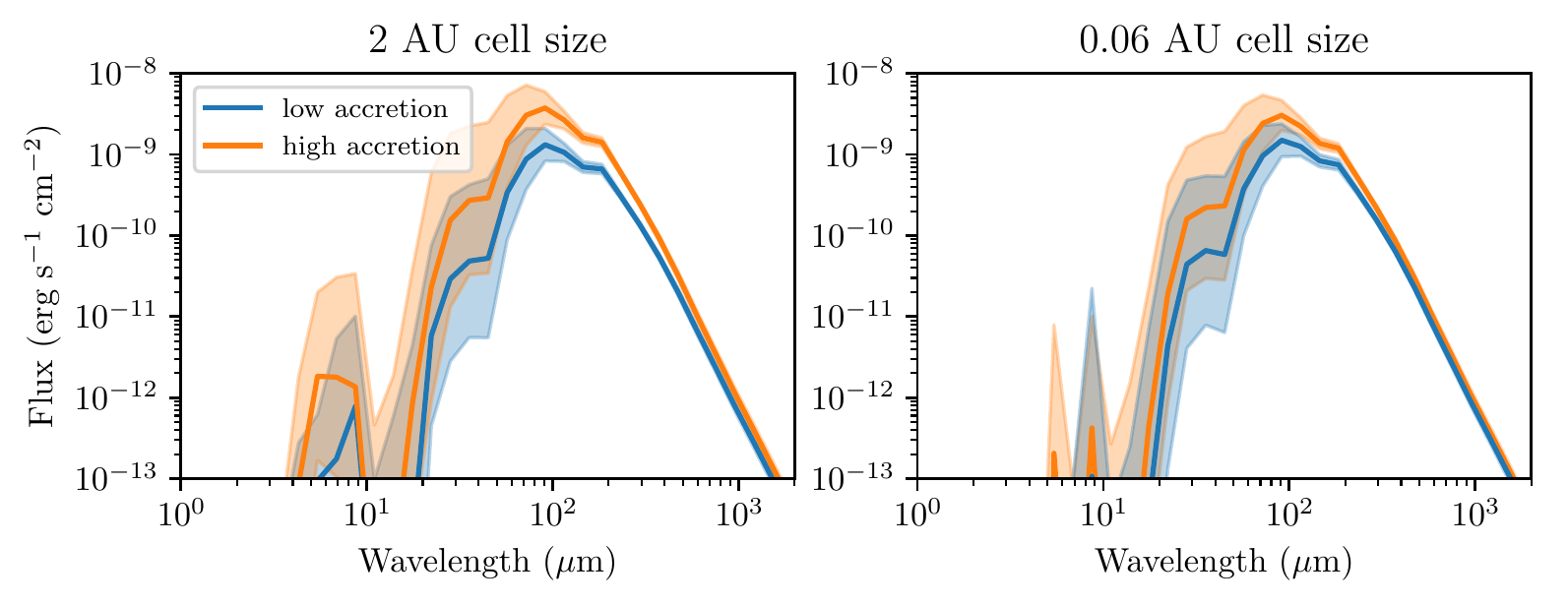} }
\protect\caption{\label{fig:sed} Low and high-accretion rate SEDs for both the 2\,au resolution case (left) and the 0.06\,au resolution (right). In each case the SEDs have been calculated along 12 different projection directions corresponding to the 12 vertices of an icosahedron to ensure a uniform distribution over the $4\pi$ solid angle. The solid line show the mean flux in each wavelength bin, while the shaded region gives the 1$\sigma$ uncertainty region.
}
\end{figure*}

\begin{table}
\caption{Synthetic observables}
\label{tbl:sed}
\begin{tabular}{cccccc}
\hline\hline
Model & Age & $L_\mathrm{input}$ & $L_\mathrm{bol}$ & $T_\mathrm{bol}$ & $L_\mathrm{smm}/L_\mathrm{bol}$ \\
      & (kyr) & (L$_{\sun}$) & (L$_{\sun}$) & (K) & (\%) \\
\hline
   \begin{tabular}{@{}c@{}} 2\,au \\ low accretion \end{tabular} & 50.4 & 10.0 & $8.7 \pm  3.5$ & $42 \pm 8$ & $1.3 \pm 0.7$ \\
\hline
   \begin{tabular}{@{}c@{}} 2\,au \\ high-accretion \end{tabular} & 50.8 & 35.3 & $27.7 \pm 16.5$ & $48 \pm 13$ & $0.8 \pm 0.5$ \\
\hline
   \begin{tabular}{@{}c@{}} 0.06\,au \\ high-accretion \end{tabular} & 50.5 & 23.7 & $21.2 \pm 10.1$ & $46 \pm 10$ & $0.9 \pm 0.5$ \\
\hline
   \begin{tabular}{@{}c@{}} 0.06\,au \\ low accretion \end{tabular} & 50.6 & 10.9 & $ 9.7 \pm 3.8$ & $41 \pm 8$ & $1.3 \pm 0.7$ \\
\hline
\end{tabular}
\end{table}

We calculate radiative transfer models for four different snapshots; two for each available resolution with one model corresponding to a low accretion rate and the other to a high-accretion rate. \Figure{sed} shows the SEDs calculated for the four radiative transfer models, while \Table{sed} lists some basic quantities calculated from the SEDs. As described in Section~\ref{sec:radmc} the synthetic SEDs are calculated along 12 different lines of sight distributed evenly over the $4\pi$ solid angle, and the quantities listed in \Table{sed} are the average and 1sigma standard deviations calculated from the individual SEDs.

As can be seen from \Table{sed} the average bolometric luminosity inferred by integrating over the SEDs is always somewhat smaller than the input luminosity; something that can be attributed to the limited size of the aperture, which means that some fraction of the emission from the system remains uncovered. Also, the uncertainty on the bolometric luminosity is quite large, which indicates that the inferred luminosity depend strongly on the line-of-sight. This is a well-known issue for systems that contain a disc, where shielding in the disc plane can easily attenuate the emission from the central by a factor of 2 (e.g. \citealt{Whitney:2003ke,Frimann:2016dy}). The situation is further aggravated in the situation at hand since the system is traversed by a massive filament running roughly perpendicular to the disc, meaning that many lines of sight close to face-on to the disc are also subject to high column densities.

\Table{sed} also shows the bolometric temperature, $T_\mathrm{bol}$ \citep{Myers:1993en} and the ratio between the sub-millimetre luminosity (the integral of the SED over wavelengths >\,350\,$\mu$m) and bolometric luminosity, $L_\mathrm{smm}/L_\mathrm{bol}$ \citep{Andre:1993cz}, both of which are commonly used tracers of the evolutionary stage of the protostar. The low value of $T_\mathrm{bol}$ (<70\,K) and high value of $L_\mathrm{smm}/L_\mathrm{bol}$ (>0.5\,\%) shows that observationally the protostellar system would be consistent with being a deeply embedded Class~0 source. The embedded nature of the source can also be appreciated by looking at the SEDs in \Fig{sed}. The emission is hardly detected at wavelengths <10\,$\mu$m, and the width of the shaded regions show that projection effects play an important role indicating that the dust is optically thick. At wavelengths $\gtrsim$100\,$\mu$m when the dust becomes optically thin the emission from the SED becomes independent of the line-of-sight.

For the 2\,au resolution case the input luminosity increases from 10.0 to~35.3\,$L_\odot$ between the two snapshots corresponding to a fractional increase of a factor 3.5. We can also estimate the observed fractional increase by integrating over the SEDs and calculating the ratio for each of the 12 lines of sight. Doing this yields a fractional luminosity increase of 3.1\,$\pm$\,0.4\,$L_\odot$. For the 0.06\,au resolution case the fractional increase in the input luminosity is somewhat smaller at a factor of 2.2, while the observed fractional increase is 1.9\,$\pm$\,0.4\,$L_\odot$. Looking at the SEDs in \Fig{sed}, we see that high-accretion SEDs have systematically greater fluxes relative to the low accretion SEDs for wavelengths $\gtrsim$\,20\,$\mu$m. Below 20\,$\mu$m the SEDs are dominated by noise because of the extremely high optical depths inherent to embedded protostars. The fractional flux increase is not uniform over the SED but is greater at shorter wavelengths, and consequently bursts will be easiest to detect at the shortest wavelength where the source can be detected. For the 2\,au resolution case the fractional increase of the flux at 25, 70, 100, and 1000\,$\mu$m is 7.4, 4.1, 2.8, and 1.4. For the 0.06\,au resolution case the fractional increases at the same wavelengths are 5.5, 2.6, 2.0, and 1.2.

The fractional luminosity increases studied here are smaller than classical FUor outburst, which typically increase their luminosities by a factor of 10 or more. However, the fractional luminosity increases is significantly above the typical variability observed in young stars, which is typically below 20\% at mid-infrared wavelengths (e.g.\ \citealt{Rebull:2015kn,Flaherty:2016bt,Rigon:2017hy}).

\section{Discussion}
\subsection{Implications of late infall}
The intermittently high-accretion rate at late times for sink b harboured in a high-density environment indicate that star--disc systems can be fed by gas
that initially was more than $10^4$ au away from the star at the time of stellar birth.
Our small sample does not allow for a statistical study of the frequency of these late infall events.
However, the weaker signs for another case
(sink 8 in K17, not discussed in this work) as well as significant mass increase of several sinks after $t>100$ kyr
in our parental run indicates that such infall is common.
As also discussed by \citet{2014ApJ...797...32P}, we suggest that late infall events is an important part of the solution
to the luminosity problem encountered for YSOs.
Late infall may be the reason for recent observations of
misaligned outer and inner discs \citep{2016ApJ...830L..16B,2017A&A...597A..42B}.
In the case of the binary system IRS 43, \citet{2016ApJ...830L..16B} found that the circumbinary disc
has a different orientation than the orbital plane of the binary, as well as their corresponding discs.
\citet{2017A&A...597A..42B} find a misalignment of 72 degrees between the inner and outer disc around HD 100453.
Considering the underlying turbulent motions in the molecular cloud, gas that falls on to the system at later times
is likely to have a different angular momentum than the net angular momentum of the star--disc system
at the time of the infall.
If the rotational component is strong enough, the system may well establish an outer misaligned disc.
Moreover, infall of material with different angular momentum may also lead to changes of the disc properties 
such as the disc size and disc mass \citep{2015A&A...573A...5V}.

Furthermore, late infall may have profound consequences for the formation of planets.
Typical disc lifetimes of about 1 to a few Myr \citep{2001ApJ...553L.153H,2009AIPC.1158....3M}
and the observations of gap and ring structures in young discs \citep[e.g.][]{2015ApJ...808L...3A}
indicate that planet formation starts early after star formation.
Potentially, late infall can thus also enrich the mass reservoir in the disc at times during planet formation.
While outside the scope of the current paper, we plan to continue the simulations, in particular the ones for the
more embedded sinks (sinks b and d), to constrain the potential importance of late infall at times much later than 100 kyr.

While it is certainly possible that fragments landing directly on the stellar surface could be part of the solution, 
our results suggest a scenario of infall and disc instabilities as the explanation for episodic accretion.
The current selection of zoom-in runs were chosen to be in relatively quiet areas of the star-forming region. 
Looking at the selection of solar mass stars in \citet{jensen} much more violent infall rates are possible. 
There are also indications \citep{2017A&A...602A.120F} that episodic accretion may be linked with binary systems.

\subsection{Limitations in the zoom-simulations}
Carrying out deep AMR simulations covering all scales from a GMC down to protoplanetary disc scales
is technically challenging and computationally expensive, even if done for a single star.
We have made many changes compared to the public version of \ramses, which collectively have increased both
the single core performance and the scalability of the code. Nonetheless,
we cannot currently afford to include all physical processes potentially relevant in protoplanetary discs to full extent, nor can we
apply arbitrary high-resolution for simulating several thousand yr of disc formation and evolution yet.
Recent developments of new codes (e.g. \athenaplus\footnote{\url{http://princetonuniversity.github.io/athena/}})
or code frameworks such as \dispatch\ \citep{2017arXiv170510774N} will help to overcome some of these limitations.

Magnetic fields are a crucial component in star formation,
and we account for the magnetic fields present in GMCs during the formation and evolution of
protoplanetary discs consistently.
Together with the underlying differences in the morphology of the gaseous filaments in the cloud,
we are thus able to naturally explore different conditions for star and protoplanetary disc formation.
However, we assume the gas to be perfectly coupled to the magnetic fields and use ideal MHD in our simulations.
It is well known that the gas in protostellar systems is far from being perfectly coupled, due to very low degrees of ionizations,
and non-ideal MHD effects such as Ohmic dissipation, ambipolar diffusion, or the Hall effect can become important during the formation
\citep[e.g.][]{2008ApJ...676.1088M,2012A&A...543A.128J,2013A&A...554A..17J,2011ApJ...738..180L,2016MNRAS.457.1037W,2016ApJ...830L...8H,2016A&A...587A..32M} and evolution of stars and protoplanetary discs \citep[e.g.][]{2015ApJ...801...84G,2017A&A...600A..75B,2017ApJ...836...46B}.
These effects would generally cause a decreasing magnetic field strength, and
we most likely overestimate the effects of magnetic braking,
as well as the strength of magnetic fields inside our discs.
As discussed in K17, we expect to underestimate the presence, size and mass of the discs in our simulations partly
because of the ideal MHD approach.
Considering the potential overestimate of the magnetic field strength,
the magnetic field further prevents gravitational instabilities inside the discs in our simulations.
It is beyond the scope of this work to study the details of the magnetic field structure.
We refer the reader to \citet{2017arXiv171011195K},
in which we investigate the strength and topology of magnetic fields in discs inherited from and anchored in the fields penetrating the
GMC.

The other relevant pressure component in our disc is thermal pressure.
\citet{2015MNRAS.446.1175T} demonstrated that radiation transfer causes heat-up of the disk even in its outer parts.
We have not included radiative transfer \citep[e.g.~by using the ray-tracing scheme implemented
in the Copenhagen version of \ramses][]{2015MNRAS.453.1324B}, and since our table based cooling implementation
assumes radiatively thin cooling above a temperature floor of 10 K, our discs are generally colder than they would be in reality.
A hotter disc would be more stable, with an additional thermal pressure, suppressing the occurrence of gravitational instabilities in the disc,
in contrast to the effect of a lower magnetic field.
Furthermore, the infall of clumps on to the star--disc system causes a local heat-up inside the disc.
In principle, if the heat-up is efficient enough, it may hinder the onset of a gravitational instability in the disc.
In reality however, we expect the heating to be not so drastic
since mass accretion scales to the temperature as $\dot{M} \propto T^4$.
Moreover, gravitational instability in the disc may occur for short cooling times,
whereas longer cooling times suppress it \citep{2001ApJ...553..174G,2004MNRAS.351..630L,2005ApJ...619.1098M,2012MNRAS.427.2022M,2012MNRAS.421.3286P,2015ApJ...814..155B}.

The accretion rates in the low and the high-resolution models of sink b are on average consistent though the presence of gravitational instabilities 
in the low resolution run induces bursts lasting for $\sim 100$ yr. 
As explained above the gravitational instabilities in the low resolution run at this time are induced by a mass buffering effect.
In the high-resolution run we resolve the inner parts of the disc, which circumvents the pile up of gas at $\approx 10$ au. 
Therefore, gravitational instabilities are absent in the high-resolution run around $t=50$ kyr,
and hence the accretion rate varies less in the high-resolution run than the low resolution run at a cadence of 10 yr. 
Comparing the accretion profile sampling with a cadence of 10 yr to that sampled with 1 yr, closer to
the orbital time at distances < 1 au from the sink particle, we find larger fluctuations in the high-resolution run.
In the case of the low resolution run 10 yr correspond to about two orbital times 
at the resolution of 2 au for the mass of sink b of $m\approx0.3$ M$_\odot$ at $t\approx 50$ kyr.
Considering the the minimum cell size of $0.06$ au as a radius, 10 yr correspond to more than 365 orbital times.
Given that the sink accretes mass mostly from the innermost cells, plotting with a 10 yr cadence
we partly average out the resolved fluctuations present in the high-resolution run by using a lower cadence. This can be
appreciated in \Fig{acc27_prof}, where the 1 yr cadence is shown by a thin green line.

In \Fig{part_acc_angle_distribution}, we illustrated the accretion angle of the gas on to the sink.
However, at the current limits of resolution we do not capture adequately the real paths taken
by gas accreting to the protostellar surface. The protostar itself is magnetized and depending on the accretion rates
gas may accrete on to it in funnels at distances
much smaller than 1 au in the form of magneto-spheric accretion \citep{1994ApJ...426..669H}, or through a magnetized
boundary layer.
To constrain the thermal efficiency parameter $\alpha$ precisely based on our zoom-simulations is
therefore not possible, due to the combination of lacking resolution and microphysics.
Nevertheless, the simulations show that gas can accrete on to the star at a range of angles,
but if a disc is present the gas preferentially approaches the protostar along the disc mid-plane.
The existence of even a small circumstellar disc may therefore be an important factor in regulating the thermal
efficiency of the accretion flow. 

\subsection{Protostellar evolution}
We have calculated the protostellar evolution assuming a dynamic model of the thermal efficiency near the accretion shock.
The evolution of the protostellar structure depends to some extent on the exact model used \citep{jensen}.
The radial expansion during the initial main accretion phase and during episodic burst later on depends on the upper limit on $\alpha$.
A similar effect is not present for the effective temperature of the protostar,
which is much more insensitive to the details of the thermal efficiency $\alpha$.
This is partly a result of the way accretion energy is distributed in the model. In our implementation the specific accretion energy is
distributed evenly across the outermost radial shells of the protostar as opposed to depositing the entire energy in the outermost shell.
We set an upper limit on the specific energy added to each radial shell, which regulates the extent of the region where the energy is deposited.
During a burst the accretion energy will therefore be spread out over a larger amount of radial shells rather than resulting in an increased
energy injection in the outermost cell. This effect can cause the effective temperature of the protostellar models to be underestimated,
but it should not significantly alter the evolution of the protostar, since in any case convection will rapidly redistribute the added
heat \citep{jensen}.

We expect models with a higher thermal efficiency to have longer contraction times, which delay the onset of deuterium burning
as well as the overall evolution of the models \citep{2017A&A...599A..49K}.
However, this effect is negligible in our study, where we are mostly interested in significant luminosity enhancements
induced by episodic accretion at early times. In principle, an accretion burst, such as those observed for sink b, both changes the
luminosity and the effective temperature. It therefore correspond to a diagonal displacement in the Hertzsprung--Russel diagram.
This is important for more evolved protostars that are not embedded anymore, where the change in temperature can result in an
overestimate of the mass. In the embedded case though changes in the stellar SED is masked, since all radiation is reprocessed by the dust.

\subsection{Comparison with observations}

OO~Serpentis is a Class~I protostar in the Serpens NW star-forming region, which was observed to undergo an accretion burst in 1995. \citet{2007A&A...470..211K} found ratios of the peak brightness relative to the preoutburst brightness >\,10 at wavelengths 2.2, 12, 25, and 60\,$\mu$m. In particular, the fractional increase at 25\,$\mu$m was measured to $\approx$25. The detection at wavelengths <10\,$\mu$m in OO~Serpentis is consistent with OO~Serpentis being a Class~I object, whereas the synthetic observations point towards the simulated source being a deeply embedded Class~0 object.
Probably, the best known direct observation of an accretion burst in a deeply embedded object is the case of HOPS~383,
a Class~0 protostar in Orion, which was observed to increase its 24\,$\mu$m flux by a factor of 35 between
2004 and 2008 corresponding to an increase in luminosity from $\sim$0.2 to $\sim$7.5\,L$_{\sun}$ \citep{Safron:2015ew}. The fractional increase of the brightness at 24--25\,$\mu$m of OO~Serpentis and HOPS~383 is significantly larger than the corresponding increase of a factor of 7.4 and 5.5 seen for the 2\,au and 0.06\,au resolution burst.
Recently, \citet{2017ApJ...837L..29H} detected a burst in the massive protocluster NGC~6334I quadrupling its brightness at mm wavelengths corresponding to a fractional increase of its luminosity of $70\pm20$. We remark that NGC~6334I is not a single object but a massive cluster consisting of $\sim$5 individual sources, and thus the situation may not be completely equivalent to that of a low-mass single protostar like the one studied here.

The simulated bursts discussed in this work are much weaker indicating that such drastic bursts as discussed above are exceptional. Nevertheless, we stress that although the increase in luminosity of the simulated bursts is smaller than the above observations or the order of magnitude increases of a classical FUor type object, they are still significant enough to be observable. For example \citet{Scholz:2013ij} surveyed 4000 YSOs in the Galactic plane over 5\,yr and found only four with a fractional flux increase >2.5 at wavelengths of 3.6\,$\mu$m, and 4.5\,$\mu$m, which is comparable to our simulated bursts.
One problem to constrain the frequency of episodic accretion events is the difficulty in determining
whether a protostar underwent an accretion burst event in the past.
A promising method to circumvent this issue is to look for extended C$^{18}$O emission as a signpost of previous
accretion bursts \citep{Jorgensen:2015kz,2017A&A...602A.120F}.

\section{Conclusion}
In this work, we investigated the accretion process of six young protostars that are formed in a
Giant Molecular Cloud. We followed the stars during up to 180 kyr of evolution after the protostellar birth.
By applying adaptive mesh refinement, we are -- for the first time --
able to follow the formation and evolution of star--disc systems
accounting both for gas infall and disc instabilities consistently.
The environment around the protostars is determining for the global evolution in the accretion profile, 
and we find significant variations among the protostars in the overall shape of both their early and late accretion histories.
In particular, we find that the presence of a disc increases the frequency of accretion bursts, while the amplitude is related
the mass reservoir in the disc.   
We find that infalling gas acts as the main supplier for the disc and thus for the occurrence of high-accretion rates
also at times when the protostar is otherwise already evolved.
A detailed analysis of the accretion process shows that mass may accrete on to the sink with a significant angle to the disc midplane.
Specifically, a run carried out with higher resolution shows that mass blobs may fall 
vertically on to the inner part of the disc. 

In simulations with a maximum resolution of 2 au,
local gravitational instabilities in the disc occur around 20 -- 40 au from the sink
that correlate with short accretion bursts up to $\sim 10$ higher than prior to the burst on time-scales of about $100$ yr
in agreement with previous results by \citet{2011ApJ...730...32S,2012MNRAS.427.1182S,2012ApJ...747...52D,2015ApJ...805..115V}.
Follow-up simulations with a maximum resolution of 0.06 au in the vicinity of the sink
show similar overall accretion profiles, but lack the existence of gravitational instabilities in the disc,
and hence do not show as accentuated accretion bursts as those seen in the lower resolution run, though the
intermittency goes up as a result of the shorter orbital time-scale resolved in the high-resolution run.
We find that the gravitational instabilities in the low resolution run are induced
by a buffering effect probably due to insufficient resolution for the inner disc. Due to the computational difficulty of
following the disc evolution with a maximum resolution of 0.06 au for more than 1000 yr, which is a short period
compared to the disc lifetime, we do not know
if the high-resolution run will evolve similar disc instabilities as the low resolution run at later times.

By looking at the mass evolution around the protostars,
we find that for the six sinks we have studied the gas associated with the prestellar core collapses within $\sim$10 kyr to a
more compact object with an envelope of $\approx$1000 au in radius. Most of the selected protostars have a relatively
uncomplicated accretion history, due to isolated formation sites in the GMC, which from a numerical point of view make them
easier to evolve. In a few cases though -- sink b being the most prominent -- the protostars are located in a more
cluster like environment, where the envelope at later times can be fed typically by local high-density filaments.
Late infall of gas with an angular momentum orientation different from the initial envelope and
the already formed star--disc system may be the explanation for recent observations
of systems with misaligned inner and outer discs.

To properly constrain the effects of episodic accretion on the protostellar evolution,
we studied the accretion process in more detail by evolving the characteristics of the protostars
with the stellar evolution code \mesa.
We find that the accretion luminosity corresponding to the rapid enhancement of the protostellar accretion rate
triggers luminosity bursts.
As an additional consequence of such accretion events, the temperature and radius of the protostar increases.
To compare directly with observations,
we post-processed the simulations with the radiative transfer tool \radmc\
to study the resulting spectral energy distributions of the protostars in detail.
In line with previous works, bursts in the accretion rate enhance the flux of the protostar
at sub-mm and shorter wavelengths.
In particular, we study one burst period in more detail and find enhancements of a factor of about 5 at 24 $\mu$m wavelength
Such enhancements at this wavelength, even though lower than the observed increase of a factor of 35 for HOPS 383 \citep{Safron:2015ew} would be possible to observe.

\section*{Acknowledgements}
We thank {\AA}ke Nordlund for his contributions to the development of the zoom-technique and the modified \ramses\ version 
used at the Centre for Star and Planet Formation in Copenhagen, and for valuable suggestions and comments to the article.   
We thank the developers of the python-based analysing tool {{\sc yt}} 
\url{http://yt-project.org/} \citep{2011ApJS..192....9T}. 
Their efforts in providing analysis support for \ramses\ simplified and improved our analysis significantly. 
We are grateful to the anonymous referee for constructive feedback that helped to improve 
the manuscript.
This research was supported by a Sapere Aude Starting Grant from the Danish Council for Independent Research to TH.
Research at Centre for Star and Planet Formation is funded by the Danish National Research Foundation (DNRF97).
We acknowledge PRACE for awarding us access to the computing resource CURIE based in France at CEA for
carrying out part of the simulations.
Archival storage and computing nodes at the University of Copenhagen HPC centre, funded with a research
grant (VKR023406) from Villum Fonden, were used for carrying out part of the simulations and the post-processing.

\bibliographystyle{mnras}
\bibliography{refs} 

\appendix
\section{Numerical Convergence}
\begin{figure}
\subfigure{\includegraphics[width=\linewidth]{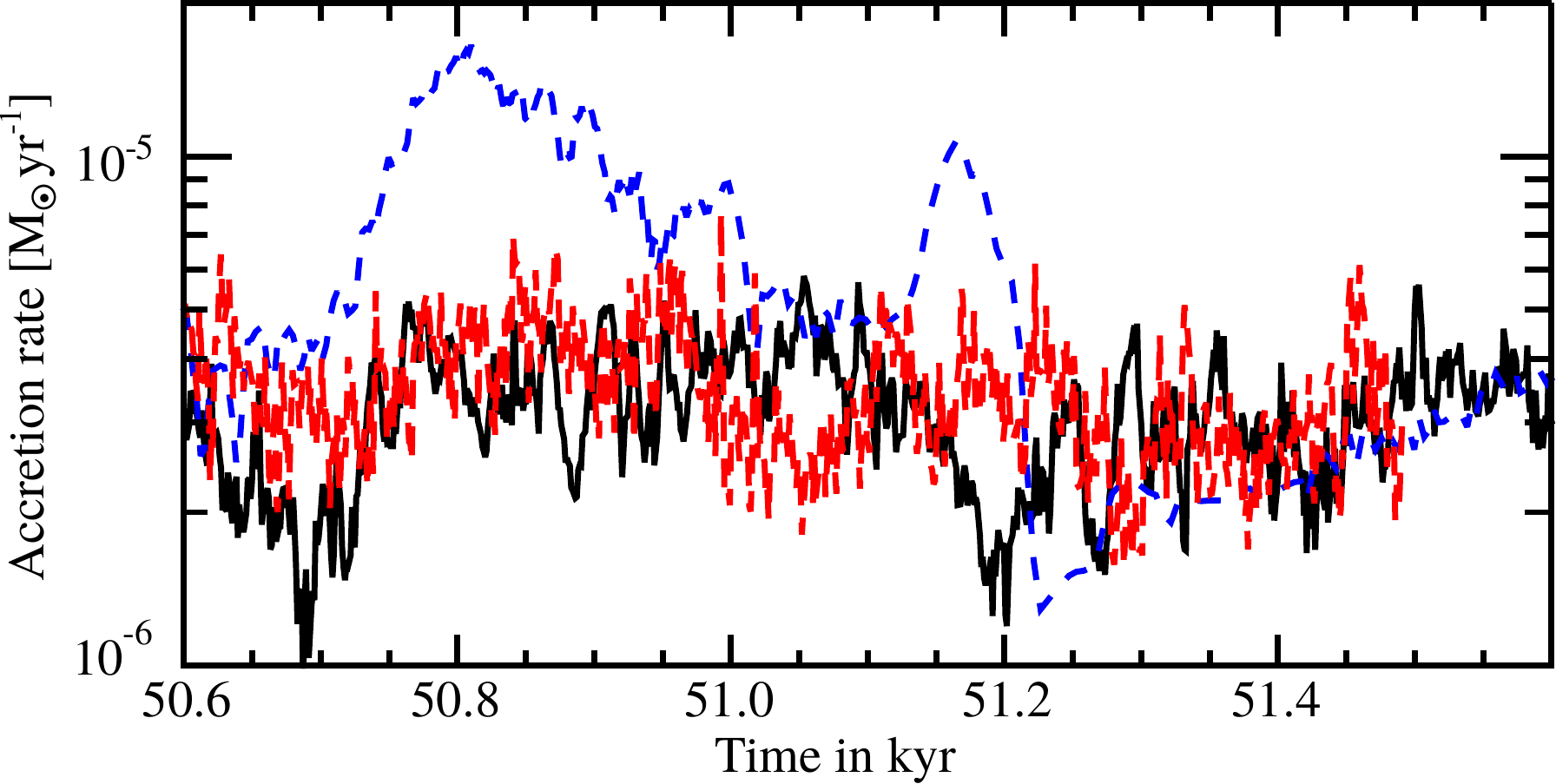} }
\subfigure{\includegraphics[width=\linewidth]{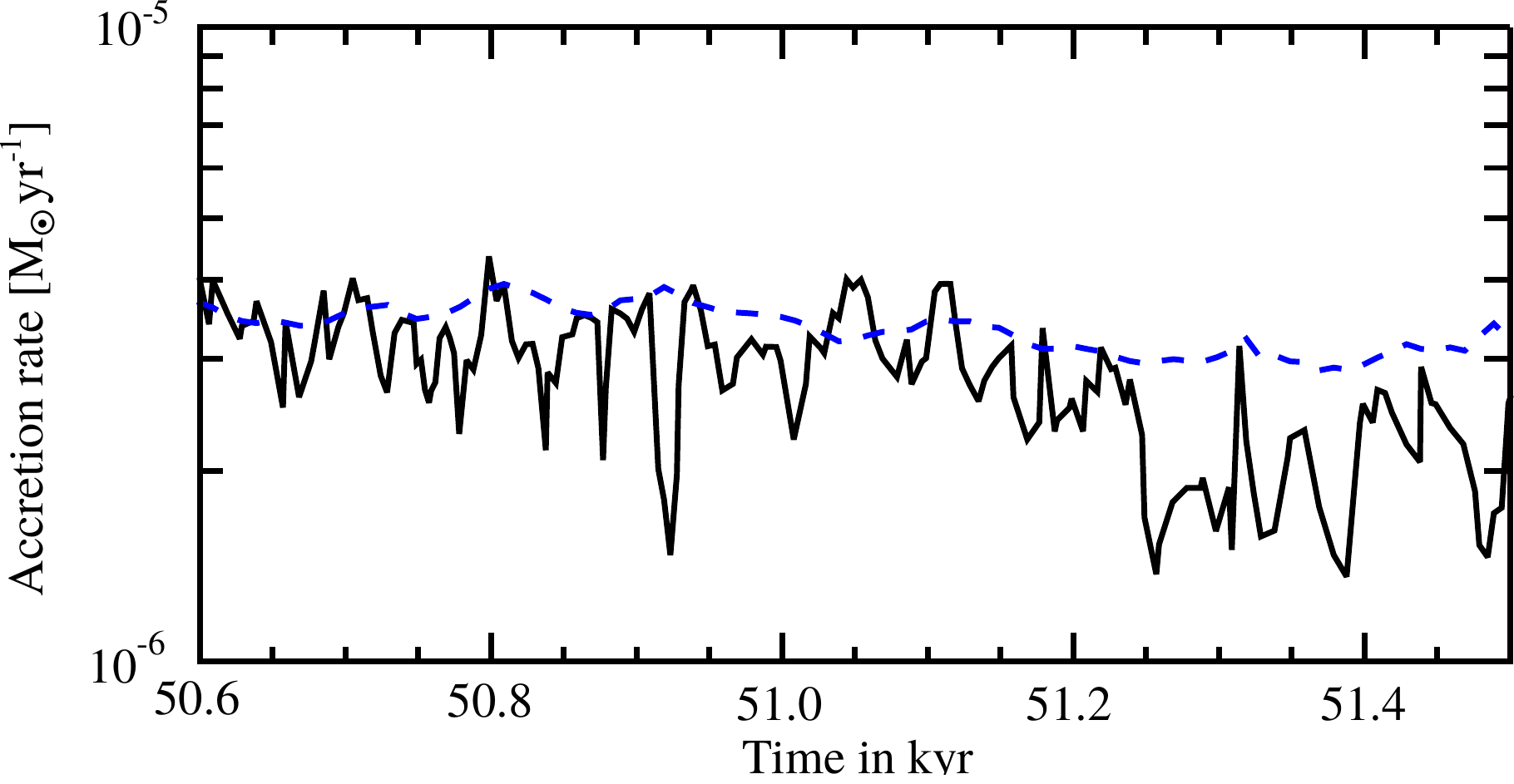} }
\subfigure{\includegraphics[width=\linewidth]{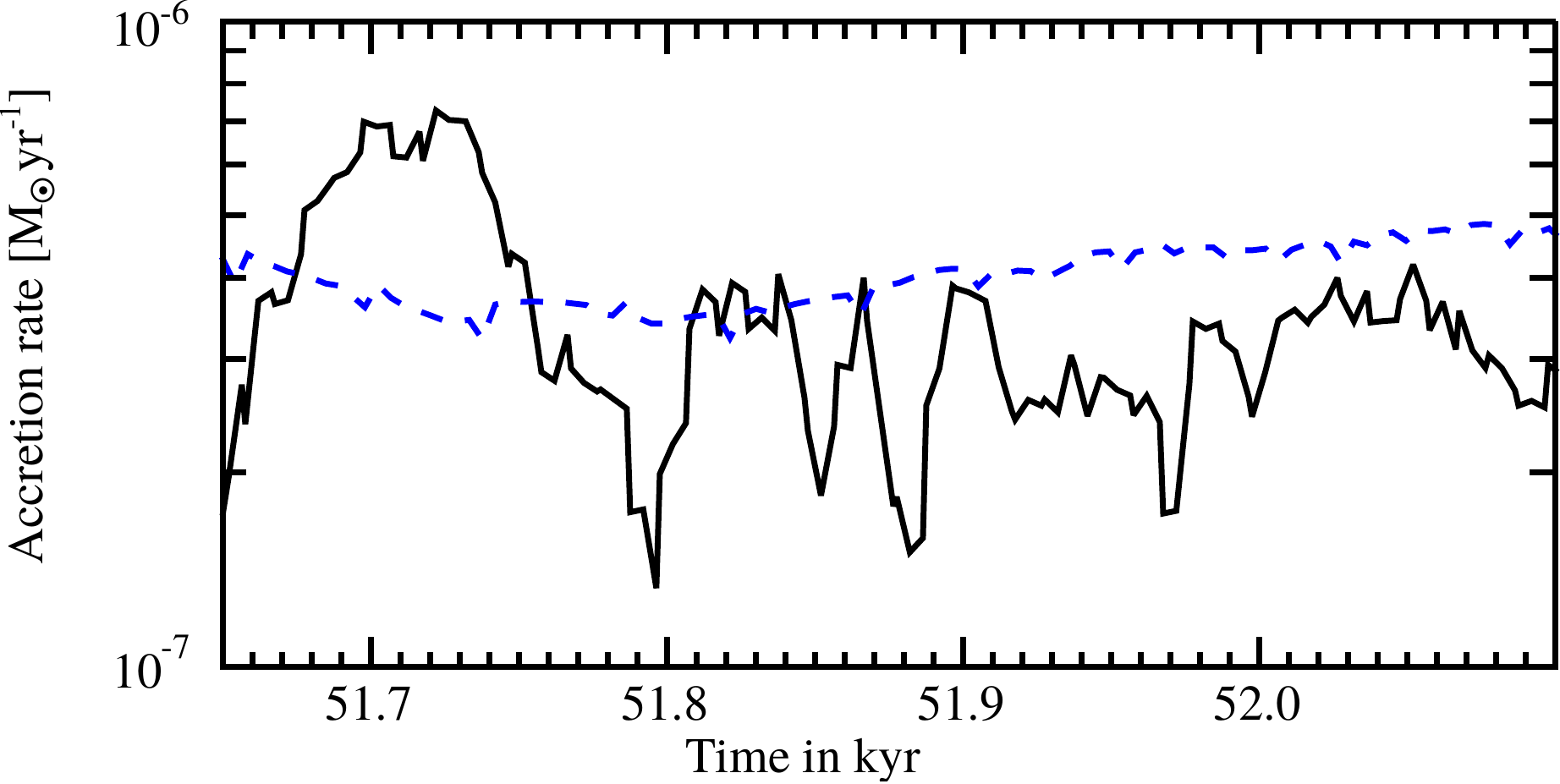} }
\protect\caption{\label{fig:acc_comp_prof} Accretion rates for sink b (top), sink d (middle) and sink f (bottom) 
of runs with 2 au resolution 
(blue dashed line) and of runs with $0.25$ au resolution run (black solid line). 
For sink b, we also plot the accretion profile of the $0.06$ au resolution run (red dotted line). 
}
\end{figure}

As discussed in the article, some of the most prominent accretion bursts in the accretion profile of the low resolution run of sink b 
are triggered by gravitational instabilities in the disc.  However, these instabilities, and hence the corresponding accretion bursts,
happen close to the accretion radius of the sink particle and seem to be induced by a mass buffering effect in the low resolution run. 
When increasing the resolution, and therefore decreasing the physical size of the accretion radius in the high-resolution run,
the effect disappears, and the amplitude of the bursts is decreased by up to a factor of 5. To investigate if this is a common trend,
and in order to constrain the numerical convergence of the accretion rates further, 
we have carried out comparison runs with a maximum resolution of $0.25$ au for sink b,
as well as for two other sinks that also maintain a circumstellar disk (sink d and sink f). In all cases we have carried out these
tests 50 kyr after star formation over a time interval of about 1000 yr.
The accretion profiles are shown in \Fig{acc_comp_prof}. 
We find that the $0.25$ au run (black solid line) agrees well with the $0.06$ au resolution run (red dotted line) 
in the case of sink b, while the $2$ au resolution run clearly differs. 
The reason for this is that we avoid the artificial mass buffering effect 
in both the $0.25$ au and the $0.06$ au resolution runs. 
Nevertheless, all three profiles are in approximate agreement, when integrating for longer time intervals.
For the two other sinks, on the other hand, there are no indications of problems.
In both cases there is good agreement for the accretion profiles between the  2 au resolution
and the $0.25$ au comparison runs. In both cases the higher resolution run has a more intermittent accretion
profile, but agrees reasonably well with the lower resolutions runs in a time-averaged sense.
Considering that we resolve the inner dynamics of the disk in more detail in the $0.25$ au resolution runs compared to the 
2 au resolution run, the lack of such fluctuations in the low resolution run is expected. 
We may see a tendency towards slightly enhanced accretion rates in the 2 au resolution runs compared 
to the higher resolution runs with $0.25$ au, but this is at best minor.
We conclude that the accretion profiles of the 2 and the $0.25$ au resolution runs are in good agreement 
for the purpose of this study and accounting for the computational costs involved in running the zoom-ins. 

\bsp	
\label{lastpage}
\end{document}